\documentclass[12pt]{iopart}

\usepackage[numbers,sort&compress]{natbib}

\expandafter\let\csname equation*\endcsname\relax
\expandafter\let\csname endequation*\endcsname\relax
\usepackage{amsmath}
\usepackage{xcolor}
\usepackage{graphicx} 
\usepackage{amssymb,theorem} 
\usepackage{epsfig}
\usepackage{hyperref} 
\usepackage[english]{babel}

\begin{document}
\title[RNN analysis of single trajectories switching between anomalous diffusion states]
{
Recurrent neural network analysis
of single trajectories switching between anomalous diffusion states
}

\author{Alvaro Lanza}
\address{Department of Physics, King's College
London, WC2R 2LS, United Kingdom}
\author{Xiang Qu}
\address{Department of Physics, King's College
London, WC2R 2LS, United Kingdom}
\author{Stefano Bo}
\address{Department of Physics, King's College
London, WC2R 2LS, United Kingdom}
\ead{stefano.bo@kcl.ac.uk}

\date{\today}

\begin{abstract}
Diffusive dynamics abound in nature and have been especially studied in physical, biological, and financial systems. These dynamics are characterised by a linear growth of the mean squared displacement (MSD) with time.
Often, the conditions that give rise to simple diffusion are violated, and many systems, such as biomolecules inside cells, microswimmers, or particles in turbulent flows, undergo anomalous diffusion, featuring an MSD that grows following a power law with an exponent $\alpha$.
Precisely determining this exponent and the generalised diffusion coefficient provides valuable information on the systems under consideration, but it is a very challenging task when only a few short trajectories are available, which is common in non-equilibrium and living systems. 
Estimating the exponent becomes overwhelmingly difficult when the diffusive dynamics switches between different behaviours, characterised by different exponents $\alpha$ or diffusion coefficients $K$.
We develop a method based on recurrent neural networks that successfully estimates the anomalous diffusion exponents and generalised diffusion coefficients of individual trajectories that switch between multiple 
diffusive states. Our method returns the $\alpha$ and $K$ as a function of time and identifies the times at which the dynamics switches between different behaviours.
We showcase the method's capabilities on the dataset of the 2024 Anomalous Diffusion Challenge.
\end{abstract}

\maketitle
\section{Introduction} \label{sec:intro}
Diffusive processes govern life at the microscale~\cite{phillips2009physical}. 
Einstein provided the mathematical description for diffusion relying on the assumption that the observed increments are unbiased, statistically independent, of finite size, and stationary~\cite{einstein1905molekularkinetischen,cohen2005history}. This formalism conveniently models many systems in physics, biology, ecology and finance~\cite{gardiner1985handbook}.
The hallmark of diffusion is a linear growth of the mean squared displacement (MSD) with time, which can be seen as a consequence of the central limit theorem~\cite{klafter2011first,aghion2021moses,vilk2022unravelling}. 
For a system described by a variable $\mathbf{X}(t)$, we define ${\rm MSD}(t) 
    \equiv {\rm E} [
       |\mathbf{X}(t) -\mathbf{X}(0)|^2
    ]$ where ${\rm E}[\cdot]$ denotes the ensemble average.
The mentioned linear increase is
\begin{equation}\label{eq:msd_lin}
    {\rm MSD}(t) 
    =2d D t,
\end{equation}
where $d$ is the spatial dimensionality of the system and $D$ is the diffusion coefficient. \\
Einstein's description of diffusion was fundamental in demonstrating that matter is made of atoms by observing their indirect effects on the motion of a tracer~\cite{cohen2005history}.
His efforts to learn the physical properties of a system by observing the statistics of a tracer immersed in it were later developed into what became the field of microrheology~\cite{squires2010fluid}. 
Einstein's assumptions for diffusive processes are very well satisfied for dilute suspensions, as the pollen grains observed by Robert Brown~\cite{brown1828xxvii} and for the motion of colloids in water, free or confined by optical tweezers~\cite{jones2015optical,argun2021simulation}. However, these assumptions are violated in crowded nonequilibrium environments such as the cell interior, in the motion of microswimmers, and in many active matter and turbulent systems.
 If any of the assumptions is violated, under general conditions~\cite{chen2017anomalous,aghion2021moses,meyer2022decomposing,mandelbrot1968noah}, the mean squared displacement grows with a power-law 
 \begin{equation}\label{eq:def}
    {\rm MSD}(t) 
    \propto K  t^\alpha,
\end{equation}
where $\alpha$ is the anomalous diffusion exponent and $K$ is the generalised diffusion coefficient  with dimension $[\mbox{length}^2\,\mbox{time}^{-\alpha}]$~\cite{Klafter2005,Barkai2012, Hofling2013, Metzler2014}.
The exponent $\alpha$ discriminates between normal diffusion ($\alpha = 1$) and anomalous diffusion ($0 < \alpha < 1$ for sub-diffusion, and $1 < \alpha < 2$ for super-diffusion).\\

Determining the anomalous exponent $\alpha$ and the generalised diffusion coefficient $K$ provides crucial information on the system being studied and sheds light on the physical processes at play, which are not directly observable, as, for example, in studies of chromatin dynamics and genome search processes~\cite{lucas20143d,zhang2016first,khanna2019chromosome,gabriele2022dynamics,yesbolatova2022formulation}.
It is also of great importance to detect changes in diffusive states and to determine the time at which they occur.
Unfortunately, in most systems of interest, it is challenging to experimentally obtain many trajectories under the same conditions~\cite{Jeon2010a}, which makes it difficult to apply methods relying on ensemble averages such as  
\cite{Golding2006,Bronstein2009,Weber2010,Jeon2013,Caspi2000,Tejedor2010,Burnecki2015,Meroz2015,Makarava2011, Hinsen2016,meyer2022decomposing,Weron2019}.
This difficulty motivated the development of methods which work on an individual trajectory level
\cite{Weron2019,Meroz2015,Elf2019,Akin2016,Sikora2017c,Weron2017,Krapf2019,vilk2022classification,Kepten2013,Burnecki2015,Kepten2015,Lanoiselee2018}), which, however, struggle when only short trajectories are available, a commonplace scenario in non-equilibrium and biophysical experiments.
The importance of inferring the anomalous exponent and the generalised diffusion coefficient, combined with the difficulty of the task, called for the development of machine-learning approaches as reviewed, {\it e.g.},  in~\cite{seckler2023machine,cai2024machine}. Methods
ranging from random forests~\cite{Wagner2017,munoz2020single,Janczura2020,Loch-Olszewska2020}, gradient boosting~\cite{Janczura2020,Loch-Olszewska2020,kowalek2022boosting}, extreme learning machines~\cite{manzo2021extreme}, (unsupervised) anomaly detection \cite{munoz2021unsupervised},
feeding statistical features in a shallow network \cite{Gentili2021},
to deep-learning approaches~\cite{Bo2019,Garibo-i-Orts_2021,Argun_2021,Kabbech2024,Szarek21,seckler2022bayesian,seckler2024change,conejero2023inferring,Kowalek2019,Granik2019,Jamali2021,gajowczyk2021detection, al2022classification,PhysRevE.107.034138,  li2021wavenet, ,Han2020,firbas2023characterization,pineda2023geometric,pacheco2024effectively,feng2024reliable,verdier2021learning,qu2024semantic,zhang2025physics} were developed.
These deep-learning methods leveraged different architectures, such as Long Short-Term Memory (LSTM) \cite{Bo2019,Garibo-i-Orts_2021,Argun_2021,Kabbech2024,Szarek21,seckler2022bayesian,seckler2024change}, Bayesian deep learning based on LSTM \cite{seckler2022bayesian,seckler2024change}, 
convolutional neural networks (CNN) \cite{Kowalek2019,Granik2019,Jamali2021,qu2024semantic,gajowczyk2021detection,al2022classification,PhysRevE.107.034138,li2021wavenet}, combinations of CNN and LSTM architectures  \cite{conejero2023inferring},
deep feed-forward networks \cite{Han2020},
convolutional transformers \cite{pacheco2024effectively,feng2024reliable,requena2023inferring,firbas2023characterization}, physics-informed transformers \cite{zhang2025physics}
and graph neural networks \cite{verdier2021learning,pineda2023geometric}.
Recent studies have also developed
 methods that learn to generate stochastic processes that closely resemble anomalous diffusion ones~\cite{fernandez2024learning}.

In 2020, the Anomalous Diffusion Challenge (AnDi) gathered the community to identify the best methods for analysing anomalous diffusion from single trajectories~\cite{munoz2021objective}. Machine-learning approaches proved to be the best-performing ones on single trajectories. 
Among these, most methods were based on LSTM architectures~\cite{Lipton2015a,Hochreiter1997} inspired by \cite{Bo2019}.
The success of machine-learning methods was also confirmed by the 2024 Anomalous Diffusion Challenge (AnDi2024)~\cite{muñozgil2024,asghar2025unet3anomalousdiffusion,feng2025enhancing}, which we will discuss in detail below.\\
Some of these methods managed to tackle scenarios in which the anomalous diffusion exponent displayed a switch between two values \cite{Bo2019,Argun_2021,munoz2021objective}. However, the case where trajectories display multiple switches between different anomalous diffusion exponent $\alpha$ and/or generalised diffusion coefficients $K$ has so far proven prohibitively difficult. To illustrate this difficulty, it is instructive to see how the method proposed in Ref.~\cite{Bo2019} to deal with trajectories displaying a single change in $\alpha$ struggles with an example trajectory featuring two switches where both $\alpha$ and $K$ vary, as shown in Fig.~\ref{fig:fig3}.
Short trajectories of this kind were the focus of Andi2024~\cite{muñozgil2024}.
We participated in this challenge and we report here on how
we tackle the challenging task of inferring the anomalous diffusion exponent $\alpha$ and generalised diffusion coefficient $K$ from a short single trajectory that displays multiple transitions between different diffusive states, characterised by different $\alpha$ and  $K$.

Building on previous work~\cite{Bo2019,Argun2020,Argun_2021} we develop a method based on recurrent neural networks employing  (LSTM)~\cite{Lipton2015a,Hochreiter1997}. 
Following the approach of Ref.~\cite{qu2024semantic}, our method returns a prediction of $\alpha$ and  $K$ as a function of time. 
We also develop a neural network to detect changes between diffusive states and identify the time at which it occurred, \textit{i.e.} the changepoint (CP).
We also discuss the possibility of inferring a CP from the time-dependent estimate of $\hat{\alpha}(t)$ and $\hat{K}(t)$ using a state-of-the-art changepoint detection algorithm~\cite{li2015atp} and contrast the performances of this approach with those of a neural network designed to identify the changepoints. 
We name our method \texttt{sequandi}, benchmark its performances on the AnDi2024 dataset and make it freely available~\cite{SEQUANDI_git}. \\

In the next paragraph, we briefly summarise the scope of The Anomalous Diffusion Challenge 2024~\cite{muñozgil2024}.
In Section~\ref{sec:architecture}, we outline the architecture and the training procedure of our method. Section~\ref{sec:alpha(t)} discusses the task of inferring the time-dependent evolution of the anomalous diffusion exponent $\alpha$ and the generalised diffusion coefficient $K$, while Section~\ref{sec:cp} focuses on finding the changepoints between different types of diffusive motion. Finally, in Section~\ref{sec:andi_results}, we apply our method to the dataset of AnDi2024 and discuss its performance.  

\paragraph{The Anomalous Diffusion Challenge 2024.}
The present work was motivated by the problems and data of The Anomalous Diffusion Challenge 2024 (AnDi2024)~\cite{muñozgil2024}. AnDi2024 is a challenge where multiple research groups competed to showcase the performances of their methods for the inference of the anomalous diffusion exponent $\alpha$ and generalised diffusion coefficient $K$ from single short two-dimensional trajectories (featuring a number of time steps, $T$, between $20$ and $200$)
from a common dataset\footnote{Part of the challenge featured the analysis of microscopy videos and ensembles of trajectories, but we focus on the single-trajectory analysis.}, as the one shown in Fig.~\ref{fig:fig1}. 
A notable difference from the first edition of the challenge~\cite{munoz2021objective} is that the trajectories can feature multiple transitions between different diffusive states. A segment between two transitions can be as short as $3$ time steps.  Establishing when such transitions occur, \textit{i.e.}, identifying the changepoint (CP) times, $t_i^{\textrm{CP}}$, is one of the tasks of the challenge.\\
The AnDi2024 trajectories are generated based on five physical models of diffusion undergoing fractional Brownian motion~\cite{muñozgil2024}. 
\begin{itemize}

\item \textit{Single state:} fractional Brownian motion trajectories characterised by $\alpha$ and $K$ without changepoints. 
\item \textit{Multiple state:} trajectories switching between two or more different diffusive states, each characterised by their $\alpha$ and $K$. 
\item\textit{Transient confinement:} trajectories switch between two different $\alpha$ and $K$ states when the particles leave or enter spatially distributed compartments, a common biological scenario.
\item \textit{Dimerisation:} considers two particles that can dimerise and begin to co-diffuse or undimerise and resume independent diffusion. From the viewpoint of a single trajectory, this is also a two-state model with switching $\alpha$ and $K$.
\item\textit{Quenched trap:} trajectories switch between a free diffusion and a completely immobile state, which they transiently occupy upon encountering a trap. 
\end{itemize}
The anomalous diffusion exponent $\alpha$ varies between $0$ and $2$: $0<\alpha<2$.
The generalised diffusion coefficient 
$K$ spans a broad range $10^{-12}\leq K\leq 10^6$. Except for the immobile state in the quenched trap model, in the other models there are no restrictions within these ranges on the allowed values of $\alpha$ and $K$ between consecutive diffusive states.

We report the performances of our method on a benchmark dataset similar to the one of AnDi2024~\cite{munoz_gil_2024_benchmark}.
We choose to focus on data from AnDi2024 as it provides a comprehensive and well-curated benchmark with several advantages. It includes biologically relevant models capturing realistic diffusion behaviours (\textit{e.g.}, confinement, dimerisation). The dataset is inspired by experimental observations featuring a diverse range of diffusion regimes and transitions, notably including changes in both the diffusion coefficient 
$K$ and anomalous exponent 
$\alpha$. Additionally, the benchmark dataset provides the ground truth, which, in combination with the fact that multiple research groups have already adopted it, enables objective and quantitative evaluation of the performance.
\section{Neural network architecture and training procedure}\label{sec:architecture}

\subsection{Pre-processing.} 

When training LSTMs, it is often convenient for the training to group the input time series into blocks.
The choice of an optimal block size results from several trade-offs, such as the minimal time resolution required, and the computational and memory cost of the training.
Since the AnDi2024 data features segments with a minimum length of 3, we choose a block to include 3 increments per spatial dimension. This means that we slice the input trajectories into successive short segments containing three increments each. We infer $K$, and $\alpha$ for each of these small segments and classify them based on whether they contain a changepoint or not.
We train the networks on the increments between subsequent trajectory coordinates. For a trajectory of length $T$, this means $T-1$ increments per each of the two spatial dimensions
$\Delta x_t = x_{t+1}-x_{t}$ and $\Delta y_t = y_{t+1}-y_{t}$.
When working with neural networks, it is customary to normalise the input data (see, \textit{e.g.}, Ref.~\cite{Bo2019,Argun2020,Argun_2021}). However, a normalisation of the increments with respect to their standard deviation or their maxima can obfuscate information about the generalised diffusion coefficient $K$. Therefore, in addition to a typical standardisation, normalising the $\Delta x_t$ and $\Delta y_t$ in each trajectory together so their combined mean is zero and variance is 1, we include their standard deviation $\sigma $ in the input vector. 

The input trajectory is then reshaped into $M$ blocks of dimension 9\footnote{Note that $\sigma $ is added on after every coordinate in the block.}, 
\begin{eqnarray}\label{eq:block}
b_\tau = [\Delta x_{3\tau},\Delta y_{3\tau},\sigma ,\Delta x_{3\tau+1},\Delta y_{3\tau+1},\sigma ,\Delta x_{3\tau+2},\Delta y_{3\tau+2},\sigma ] 
\end{eqnarray}
where the subscript $\tau$ denotes the order in sequence of the blocks and $\tau=0,...,M-1$, as shown in Fig.~\ref{fig:fig2}.
The total number of blocks $M$ is equal to the $\lceil (T-1)/3 \rceil$ ceiling division. Therefore, there may not be enough increments to fill the last block.  All such ``missing" increments are padded with a dummy value, which is later ignored by the network in a procedure known as masking. 

\subsection{Architecture.} 
As discussed in the introduction, LSTMs~\cite{Hochreiter1997,Lipton2015a}, with their ability to capture long and short-term correlations, while securing stability in the training~\cite{Lipton2015}, have proven very successful in the study of anomalous diffusion \cite{Bo2019,Garibo-i-Orts_2021,Argun_2021,seckler2022bayesian,seckler2024change,Kabbech2024,Szarek21}. We base our model on the architecture proposed in Ref.~\cite{Argun_2021}, which performed very well on the first AnDi Challenge~\cite{munoz2021objective}. As in Ref.~\cite{Argun_2021}, we choose the first LSTM to have dimension 250 and the second one 50, and adjust the relevant 
hyperparameters. In~\ref{app:hyper}, we show how this choice, motivated by the existing literature, performs equally well to networks with hyperparameters fine-tuned for this specific task.
Our method, \texttt{sequandi}, features a key difference compared to~\cite{Argun_2021}: leveraging the LSTM properties, it returns as output a time series of the same length as the input (in terms of the short segments the trajectories are sliced into).
\begin{figure}[h]
	\begin{center}
		\includegraphics[width=\textwidth]{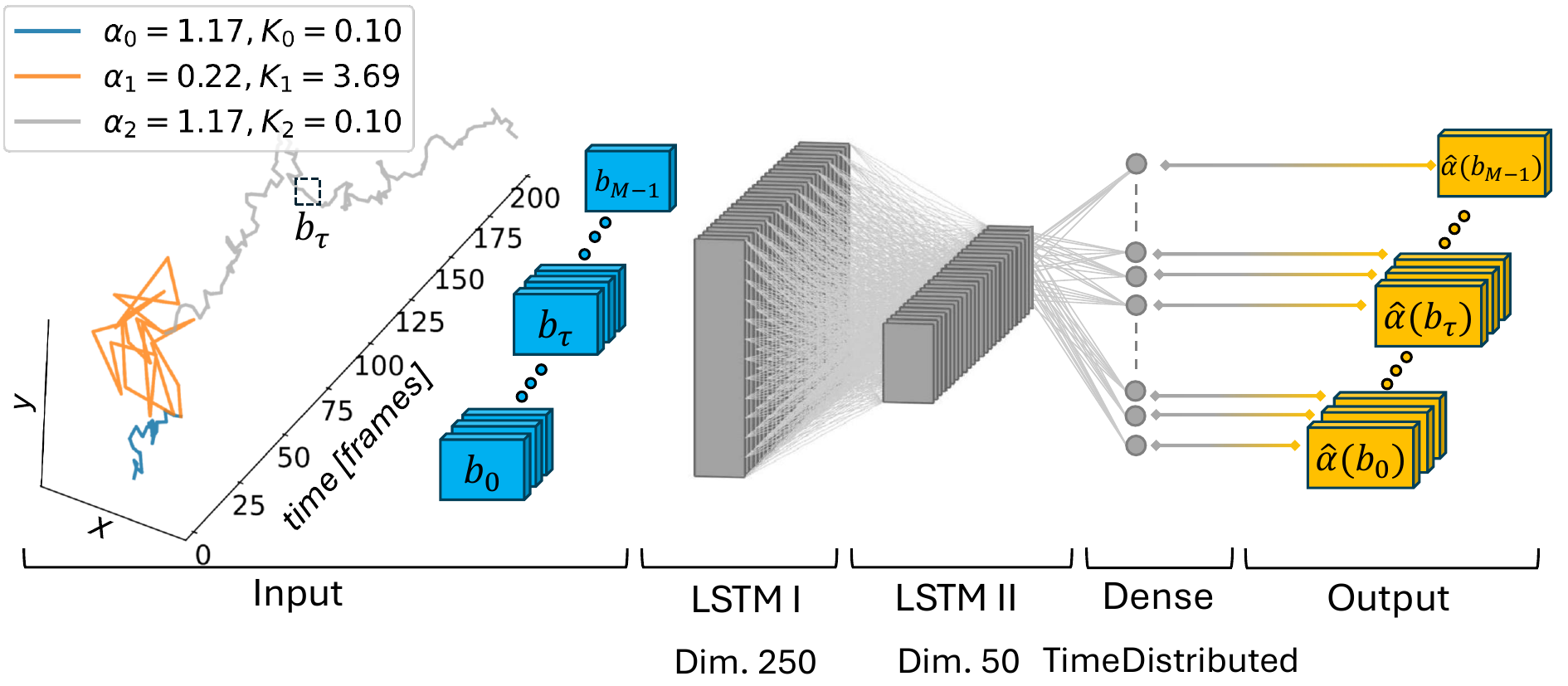}
		\caption{
        {\bf  Neural Network Architecture of \texttt{sequandi}.} 2D trajectory data is normalised and separated into $M$ blocks $b_\tau$ of 3 coordinate increments each. The blocks are then fed into two adjacent LSTM layers of dimensions 250 and 50. Next, outputs from the second LSTM layer are passed through a dense TimeDistributed layer, which is adapted depending on the task (inference of $\alpha$, $K$ or $t^{\textrm{CP}}$) and returns a sequence of length $M$, ready for post-processing.}\label{fig:fig1}
	\end{center}
	\end{figure}
This allows us to task the network to analyse trajectories of varying length and to return the time-dependent values of $\alpha$ and $K$ along the trajectory, as in Ref.~\cite{qu2024semantic}. Crucially, this key innovation makes it possible to characterise trajectories that frequently switch between different behaviours both in terms of $\alpha$ and $K$.
This would be prohibitively difficult for networks developed to study scenarios that display a single change in diffusive behaviours, such as the ones in Refs.~\cite {Bo2019,Argun_2021} (see Fig.~\ref{fig:fig3}).
In addition to inferring the time-dependent behaviour of $\alpha$ and $K$, we are interested in detecting the presence of a changepoint (CP) in each of the short segments the trajectories are sliced into, in a combined regression-classification procedure. 
Since the input trajectories have different lengths, it is not straightforward to design an architecture of variable length to classify whether any short segment contains a CP.
To this aim, we apply a fully connected layer to each of the sequential outputs returned by the second LSTM layer. This layer of variable length is known as a Time-Distributed layer and is readily available in Keras~\cite{chollet2015keras}. 
Crucially, this layer maintains the sequential nature of the LSTM layers, so that each block in the sequence is not analysed in isolation but keeps track of the correlations present in the trajectory.
Alternatively, one may have used a masking procedure, which would require specifying a maximum possible trajectory length. Finally, an output layer is attached at the end, with a specific shape depending on the task, as we will discuss below.

\subsection{Training Procedure.}

Andi2024 considered a wide range of model experiments. We therefore opted to use a training set which is mixed and contains an equal representation of each model experiment. This also opens the door to apply our network to generic diffusive trajectories, as our approach does not involve a prior screening to first ascertain the type of Challenge-specific diffusive model. 
We trained our network on simulated trajectories using the \texttt{andi-datasets} Python library provided by the challenge organisers~\cite{Munoz-Gil2020b}.
When training a network to infer a parameter from an interval, there is often a tendency to underestimate the values close to the upper limit of the interval and overestimate the ones close to the lower limit, giving rise to an S-shaped curve when plotting predictions vs the ground truth (see, \textit{e.g.},~\cite{Argun_2021}).
To mitigate this issue, we slightly biased the training dataset to have more cases with $\alpha$ closer to the extremes of the supported $(0,\,2)$ range [see Fig.~\ref{fig:app_training}(a)]. While the support for $K$ is vast, $[10^{-12}, 10^6]$, we trained for $K<35$.  A high representation of $K\leq1$ was considered as this regime is paradigmatic of confined states and has a higher relative importance in its evaluation with a logarithmic metric, discussed in the next section [see Fig.~\ref{fig:app_training}(b)]. Lastly, due to the nature of our CP detection method, we included sets of trajectories with many CP (that is, short segments). In total, we trained on $2.5\times10^6$ simulated trajectories. 

We use a batch size of 32, and a cycling of the training data of 20 epochs, with an early stopping monitor to prevent overfitting. As common with LSTM training, we apply 20\% recurrent dropout to both LSTM layers to prevent overtraining~\cite{Gal2016,Argun_2021}, and 10\% of the data is kept for validation. We use the Adam optimiser with learning rate $l_R=10^{-3}$ and numerical stability constant $\epsilon = 10^{-7}$~\cite{Kingma2015}. See Fig.~\ref{fig:app_losscurves} for the training and validation loss curves.

The particular labelling and losses for the training procedures of $\alpha$ and $K$ are discussed in Section~\ref{sec:alpha(t)} and, for $t^{\textrm{CP}}$, in Section~\ref{sec:cp}.

LSTMs can analyse time series data of varying lengths, which is of great importance for experimental applications and for studying the AnDi2024 dataset, which contains trajectories of length between 20 and 200. However, for convenience, we train the network on batches where all input trajectories are modified to have the same length $T=200$. 
As for the blocks, this modification is achieved by padding the missing values with a value, which is later ignored by a masking layer.

\section{Inferring the time evolution of the anomalous exponent $\alpha(t)$ and generalised diffusion coefficient $K(t)$}\label{sec:alpha(t)}

In this Section, we focus on the task of inferring the time-dependent (point-wise) anomalous exponent $\alpha$ and generalised diffusion coefficient $K$. We refer to our estimates as   $\hat\alpha(t)$ and $\hat{K}(t)$, while $\alpha(t)$ and $K(t)$ denote the ground truth.
As in AnDi2024, we consider trajectories that switch between different $\alpha$ and $K$, so that the ground truth $\alpha(t)$ and $K(t)$ is a piecewise constant function of $t$, as shown by the blue curve in Fig.~\ref{fig:fig3}.

\subsection{Loss function and last layer for inferring $\alpha(t)$ and $K(t)$}
The architecture of the first two LSTM layers is the same for all networks we train.
The last layer, instead, is tailored to the specific task. 
For the inference of $\alpha(t)$ and $K(t)$, the TimeDistributed output layer returns 2 floats for every block $b_\tau$ in the input sequence, yielding the block predictions $\hat{\alpha}(b_\tau)$ and $\hat{K}(b_\tau)$. The loss function for both parameters is a mean squared error (MSE) between the prediction and the ground truth. 
The ground truth of a block is defined as the time average of the $\alpha(t)$ and $K(t)$, respectively, within a block.\footnote{
For $\alpha$ (and similarly for $K$), the ground truth (label) of a block is $\alpha(b_\tau)=\frac{1}{3}[\alpha(t=3\tau)+\alpha(t=3\tau+1)+\alpha(t=3\tau+2)]$, using fewer $\alpha(t)$ if a semi-padded block.
This label is the mean of the $\alpha$ values corresponding to the increments of that particular block, allowing for a contribution from every $\alpha(t)$ in the case of a CP.}

\subsection{Post processing}\label{sec: postproc}
\begin{figure}[ht ]
    \begin{center}
		\includegraphics[width=\textwidth]{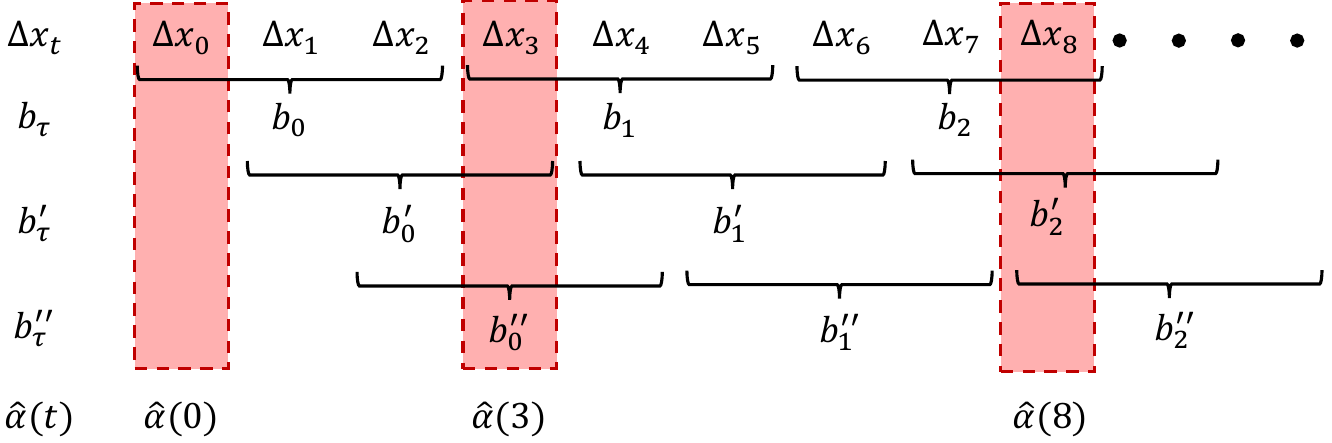}
	\end{center}
	\caption{{\bf Block assignment and point-wise inference}. The input trajectory is split into blocks $b_\tau$ containing three increments per spatial dimension as stated in Eq.~\eqref{eq:block}. 
    The network returns block-wise predictions $\hat\alpha (b_\tau), \hat K(b_\tau)$. 
    To obtain a point-wise prediction, for each increment, we shift the trajectory by one and two increments, respectively, changing which increments are assigned to each block $b'_\tau=[\Delta x_{3\tau+1},\Delta y_{3\tau+1},\sigma ,\Delta x_{3\tau+2},\Delta y_{3\tau+2},\sigma, \Delta x_{3\tau+3},\Delta y_{3\tau+3},\sigma ]$, and $b''_\tau=[\Delta x_{3\tau+2},\Delta y_{3\tau+2},\sigma ,\Delta x_{3\tau+3},\Delta y_{3\tau+3},\sigma, \Delta x_{3\tau+4},\Delta y_{3\tau+4},\sigma ]$. This returns staggered predictions $\hat\alpha (b_\tau'),\, \hat K(b_\tau'), \, \hat\alpha (b_\tau''),\, \hat K(b_\tau'')$. We obtain the point-wise $\hat\alpha(t)$ and $ \hat K(t)$ by averaging the block predictions of the blocks that contain the time $t$, illustrated by the shifting red square. 
    For instance, consider the vertical slice corresponding to $t=3$. The point-wise dynamical prediction is obtained as $\hat\alpha(t=3) = [(\hat\alpha(b_1)+\hat\alpha(b'_0)+\hat\alpha(b''_0)]/3$. For $t=0\textrm{ and }1$, we only use $\hat\alpha(b_\tau)$ predictions from blocks $\tau=0$ and $\tau=0,1$, respectively.}
		\label{fig:fig2}
\end{figure}
The block estimates $\hat\alpha(b_\tau)$ and $\hat K(b_\tau)$ are, in effect, coarse-grained predictions averaging three successive time increments. To build the refined point-wise $\hat\alpha(t)$ and $\hat K(t)$, we pool the predictions of 3 trajectories: the original trajectory, and the original trajectory shifted by one and two frames, respectively, as shown in Fig.~\ref{fig:fig2}. 
\subsection{Diffusion inference performance}\label{sec:aK_res}
\begin{figure}[h ]
    \begin{center}
		\includegraphics[width=\textwidth]{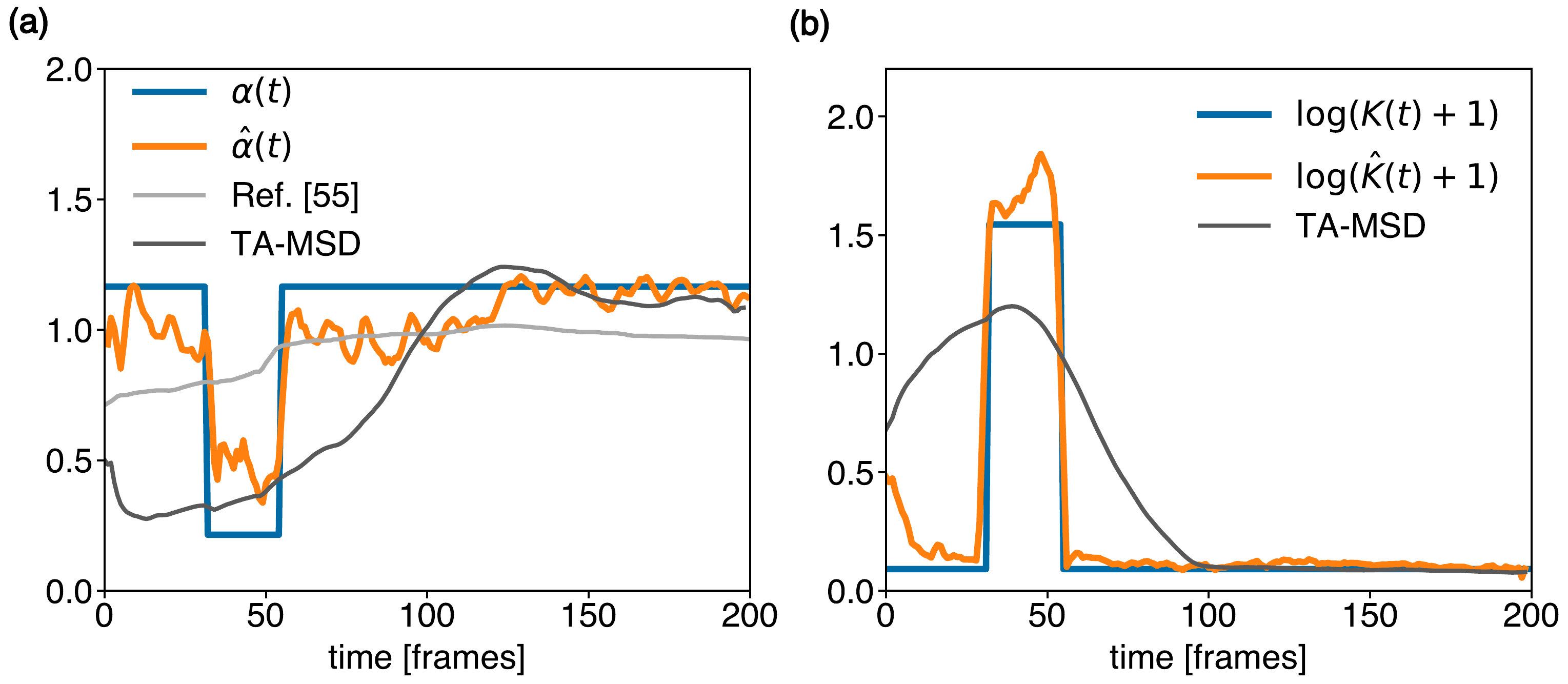}
	\end{center}
	\caption{{\bf Example of $\hat\alpha(t)$ and $\hat K(t)$ and ground truth.} Results from example trajectory in Fig.~\ref{fig:fig1}. 
	{\bf (a)} The orange line is \texttt{sequandi}'s point-wise prediction of the anomalous exponent $\hat\alpha(t)$, and the blue piece-wise constant line is the ground truth $\alpha(t)$. 
    The grey line is the prediction of the network developed in Ref.~\cite{Bo2019} and the brown line is from the TA-MSD. Since these methods require longer trajectories to work, the original trajectory was repeated (see \ref{app:compare} for details).
    {\bf (b)} The orange line is based on \texttt{sequandi}'s point-wise prediction of the generalised diffusion coefficient $\hat K(t)$, and is 
    $\log(\hat{K}(t)+1)$.
    The blue piece-wise constant line is the ground truth $\log(K(t)+1)$, and the brown line is obtained via a TA-MSD estimate. \texttt{sequandi}'s prediction tracks the ground truth with an $\textrm{MAE}_{\alpha(t)}=0.134$  and $\textrm{MSLE}_{K(t)}=0.016$. In this window of 200 frames, the error from the network of Ref.~\cite{Bo2019} is $\textrm{MAE}_{\alpha(t)}=0.268$, while for the TA-MSD they are $\textrm{MAE}_{\alpha(t)}=0.295$  and $\textrm{MSLE}_{K(t)}=0.194$
        }
	\label{fig:fig3}
\end{figure}

We analyse the performance of \texttt{sequandi} on two separate large datasets containing trajectories that were not seen by the networks during training: the benchmark dataset and the test dataset. The benchmark dataset was provided by the organisers of Andi2024, it is similar to the one used for the challenge \cite{munoz_gil_2024_benchmark}, and we discuss it in Section~\ref{sec:bench}. The test dataset is statistically similar to the dataset we used to train \texttt{sequandi}, and we discuss it in ~\ref{app:traintest}. It is mostly useful to indicate that the training did not lead to overfitting.\\
To assess the performance of our method, we compute the following metrics for the point-wise dynamic predictions $\hat\alpha(t)$ and $\hat K(t)$ in a given trajectory with $T-1$ increments 
\begin{eqnarray}
    \textrm{MAE}_{\alpha(t)} = \frac{1}{T-1}\sum_{t=0}^{T-2} \left| \alpha(t)-\hat \alpha(t) \right|, \label{eq:mae_alphat}\\ \textrm{MSLE}_{K(t)} = \frac{1}{T-1}\sum_{t=0}^{T-2}\left[ \log{\left[K(t)+1\right]} - \log{[\hat K(t)+1]} \right]^2. \label{eq:msle_Kt}
\end{eqnarray}
We measure these metrics for every trajectory in the benchmark dataset, and then take a mean over all the trajectories' errors. For this dataset, $\textrm{MAE}_{\alpha(t)} = 0.191$ and $\textrm{MSLE}_{K(t)} = 0.021$ \footnote{The AnDi2024
metrics considered a flattened array of all $t$ across all trajectories in the dataset instead of a per-trajectory average. Using this metric we obtain  $\textrm{MAE}_{\alpha(t)} = 0.176$ and $\textrm{MSLE}_{K(t)} = 0.014$.}. We discuss these performances in greater detail in Section~\ref{sec:andi_results}. For the test dataset we find $\textrm{MAE}_{\alpha(t)} = 0.191$ and $\textrm{MSLE}_{K(t)} = 0.046$.\\
To illustrate our findings, it is also instructive to visualise the predictions made by \texttt{sequandi} on example trajectories. We start by considering a trajectory from the test dataset (shown in Fig.~\ref{fig:fig1}), presenting the challenging inference of two switches, in rapid succession, between anomalous diffusion states.
In Fig.~\ref{fig:fig3}, we plot the point-wise sequential predictions. The generalised diffusion coefficient is plotted as $\log{[K(t)+1]}$\footnote{This is the argument for the mean squared logarithmic error (MSLE) used to evaluate the performance in inferring $K$ in AnDi2024 and ensures a linear-like loss for small $K$ and a logarithmic one for large $K$.}.
To visualise the improvement provided by \texttt{sequandi}, we compare it to the standard approach of computing a Time-Averaged Mean-Squared Displacement (TA-MSD) on a sliding window (brown line).
For $\alpha$ we can also compare it to an adaptation of previous state-of-the-art time-dependent inference of $\alpha$ \cite{Bo2019}, which gives the light-grey curve of Fig.~\ref{fig:fig3} (see \ref{app:compare} for details).
In~\ref{app:more}, we visualise the performance on more example trajectories.

\section{Identifying the changepoints}\label{sec:cp}

In this Section, we present our approach to detecting changes between diffusive states and inferring the time at which the changes took place. 
The trajectories we are considering often involve multiple changes and we denote by $t^{\textrm{CP}}_i$ the time at which the $i^{\textrm{th}}$ change in diffusion states occurred. 
We train a network specifically to detect and locate CPs. In Section~\ref{sec:CPalg}, we discuss the alternative approach of inferring the CP with an algorithm that studies the changes in the $\hat\alpha (t)$ and $\hat K (t)$ predictions made in Section~\ref{sec:alpha(t)}, as done, for example in Ref.~\cite{seckler2024change}.

\subsection{Output layer for changepoint detection and location}
To detect and locate changes in diffusive states, we compare the performances of two types of networks. One network is tasked to both infer the pointwise time-dependent evolution of $\alpha(t)$ and $K(t)$, \textit{and} to detect and locate changepoints.
Another network is only tasked to detect and locate the CPs, without explicitly inferring $\alpha(t)$ and $K(t)$. This second network, focusing only on the detection and location of CPs, provides the best performances.\\
The architecture of the network involves the two LSTM layers described in Section.~\ref{sec:architecture}, and a specific structure of the output TimeDistributed layer, which features \textit{two} separate TimeDistributed layers, in parallel. One layer is for detection, the other for location, specifically tasked with the following:
\begin{itemize}
    \item \textit{Detection: Is there a CP in the block?} This is a binary classification task with an integer label $t^{\textrm{CP}}_{I/O}(b_\tau)=$ 0 (in absence of CP) or 1 (in presence of CP). For this layer, we employ a standard binary cross-entropy loss function and sigmoid activation function. The model's predictions are given in one number per block, the probability that there is a CP within it. We round this number with the default threshold of 0.5 to obtain $\hat t^{\textrm{CP}}_{I/O}(b_\tau) = 0 \textrm{ or }1$.
    
    \item \textit{Location: where in the block is the CP?} Each block contains up to three different $t$ where a CP may be located, so we train this parallel output to predict on labels $t^{\textrm{CP}}_{loc}(b_\tau)=$ 0, 1, 2 or 3.
    The zero label is used when there is no CP in the block, and the non-zero labels indicate the position of the block where the ground truth CP is. 
    It is important to retain information about the order of the CPs (\textit{e.g.}, if a CP is at $t=1$, predicting a CP at $t=2$ is ``less wrong" than predicting it at $t=3$).
    We therefore choose the \textit{ordinal} categorical cross-entropy OCE~\cite{Ordinal} as the loss function. This loss function takes into consideration the order of the number of classes ($L=4$) by weighing the standard cross-entropy loss CE with the distance between predictions and ground truth
    \begin{eqnarray}
        \textrm{OCE} \left( t^{\textrm{CP}}_{loc}(b_\tau), \mathbf{\hat t^{\textrm{CP}}_{loc}}(b_\tau) \right) &&= \left( w + 1 \right)\, \textrm{CE} \left( t^{\textrm{CP}}_{loc}(b_\tau), \mathbf{\hat t^{\textrm{CP}}_{loc}}(b_\tau) \right), \\ \quad w \left( t^{\textrm{CP}}_{loc}(b_\tau), \mathbf{\hat t^{\textrm{CP}}_{loc}}(b_\tau) \right) &&= \frac{ \left| t^{\textrm{CP}}_{loc}(b_\tau) - \arg \max \mathbf{\hat t^{\textrm{CP}}_{loc}}(b_\tau) \right|}{L - 1}.
    \end{eqnarray}
    Note the predictions $\mathbf{\hat t^{\textrm{CP}}_{loc}}(b_\tau)$ are in a probability vector format (after having passed through a softmax activation function).\footnote{Unlike the definition from~\cite{Ordinal} which takes vectorised OHE labels of the $L$ classes; our equivalent implementation instead uses a \textit{sparse} CE for integer $t^{\textrm{CP}}_{loc}(b_\tau)$ labels, saving some memory and computation time.}
    Lastly, an argmax function is applied to $\mathbf{\hat t^{\textrm{CP}}_{loc}}(b_\tau)$ to predict the most likely location of the CP.
    
\end{itemize}

\subsection{Post processing} \label{sec:tCP_NN}
\begin{figure}[h!]
	\begin{center}
	\includegraphics[width=\textwidth]{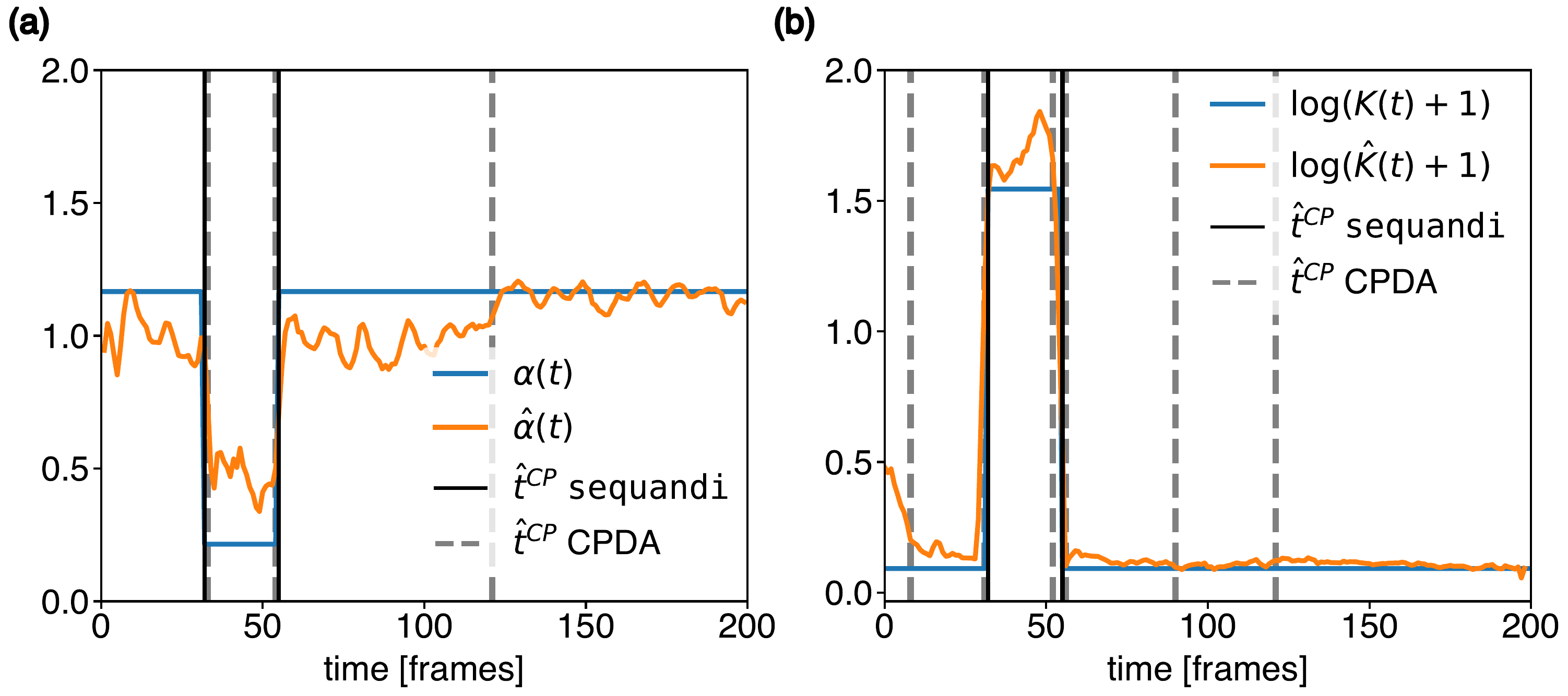}
	\end{center}
	\caption{{\bf Changepoint detection comparison: \texttt{sequandi} vs. CPDA} 
     Using the trajectory illustrated in Fig.~\ref{fig:fig1}, the two vertical solid black lines are \texttt{sequandi}'s prediction $\hat t^{\textrm{CP}}$.
     {\bf (a)} As in Fig.~\ref{fig:fig3}, the orange line is $\hat\alpha(t)$, and the blue piece-wise constant line is the ground truth $\alpha(t)$.  The three vertical dashed gray lines are obtained by the CPDA algorithm described in Section~\ref{sec:CPalg} applied to $\hat\alpha(t)$ with a confidence level of $99.99\%$. 
    {\bf (b)} As in Fig.~\ref{fig:fig3}, the orange line is 
    $\log(\hat{K}(t)+1)$ and
    the blue piece-wise constant line is the ground truth $\log(K(t)+1)$. The six vertical dashed gray lines are obtained by the CPDA algorithm described in Section~\ref{sec:CPalg} applied to $\log(\hat K(t)+1)$ with a confidence level of $99.99\%$.     
     }
	\label{fig:fig4}
\end{figure}

As for the inference of $\alpha(t)$ and $K(t)$ discussed in Section~\ref{sec:alpha(t)}, we pool the results from three versions of the same trajectory by shifting it by one and two increments (see Fig.~\ref{fig:fig2}). 
In this case, we have two types of signals for a CP from the two parallel final layers, one concerning the detection and one the location of the CP.
Sometimes, these two outputs are inconsistent (\textit{e.g.}, the detection layer may predict that there is no CP while the location layer may predict that a change occurred and at a specific time).
Our pooling consists of summing the six outputs (2 outputs for each of the three shifted trajectories), which for non-zero predictions each provide one count at the time(s) they are indicating.
The resulting $\hat t^{\textrm{CP}}(t)$ is a time-dependent curve peaking where the neural network consistently detects a CP, as shown in Fig.~\ref{fig:app_tCP}. 
We then search for all peaks that contain at least the threshold value of 2 counts (chosen to mitigate the risk of missing a CP, \textit{i.e.}, a false negative) and
assign the location of the peak as the CP prediction $\hat t^{\textrm{CP}}_{i}$. 
In Fig.~\ref{fig:fig4}, we report this prediction $\hat t^{\textrm{CP}}_{i}$ as the solid vertical lines.

\subsection{Changepoint detection and location performances}\label{sec:t_res}

In the AnDi challenge, the Jaccard similarity coeffecient $\textrm{JSC}_{\textrm{CP}}$ is used as metric for the accuracy of detecting a CP.  For a set of true positive (TP) detections\footnote{To evaluate TPs in the case of zero CPs in the ground truth (single-state trajectories), the AnDi metrics algorithm counts the correct prediction of no CPs as one TP -- the case of a true negative TN, really.} 
\begin{equation}\label{eq:JSC}
    \textrm{JSC}_{\textrm{CP}} = \dfrac{\textrm{TP}}{\textrm{TP+FN+FP}}.
\end{equation}
where FN denotes the number of false negatives, FP, the number of false positives, and $\textrm{JSC}_{\textrm{CP}}=1$ for perfect detections. The algorithm used to identify these quantities from ground truth $t^{\textrm{CP}}_i$ and predicted $\hat t_j^{\textrm{CP}}$ relies on pairing segments $i, j$ together by minimising the distance between them $d_{i,j}=  \min{\left(|t_i^{\textrm{CP}}- \hat t_j^{\textrm{CP}}|,d_{\textrm{max}}\right)}$, given a maximum pairing penalty $d_\textrm{max}$ of 10~\cite{muñozgil2024}. Often, the number of predicted CPs does not match the true number $N$, to which the Hungarian algorithm is employed to find the optimal pairings~\cite{crouse16hungarian}.\\ 
For the CPs that are paired, the precision of the inferred location, \textit{i.e.} how close to a true CP is the predicted one, is evaluated by the root mean squared error $\textrm{RMSE}_\textrm{CP}$ metric
\begin{equation}\label{eq:RMSE}
    \textrm{RMSE}_\textrm{CP} = \sqrt{\frac{1}{N_{TP}}\sum_{\textrm{paired $i, j$}}^{N_{TP}}(t_i^{\textrm{CP}}- \hat t_j^{\textrm{CP}})^2}.
\end{equation}
Importantly, the weight of the mean inside the square root is by the number of TPs $N_{TP}$ that have been found by the algorithm.
For a given set of trajectory data, one pools together all of the found TP segments across trajectories to calculate these metrics, as with the FNs and FPs in the $\textrm{JSC}_{\textrm{CP}}$.

For all the trajectories contained in the benchmark dataset, the detection metrics of \texttt{sequandi} are $\textrm{JSC}_\textrm{CP}=0.577$ and $\textrm{RMSE}_\textrm{CP}=1.504$.

\subsection{Detecting a changepoint from the inferred $\hat{\alpha}(t)$ and $\hat{K}(t)$ with an algorithm.}\label{sec:CPalg}
An alternative way of detecting and locating a CP is to exploit the pointwise time-dependent prediction of $\hat{\alpha}(t)$ and $\hat{K}(t)$, and apply a changepoint detection algorithm, as done, for instance, in Ref.~\cite{seckler2024change}. 
Here, we apply the Changepoint Detection Algorithm (CPDA) developed in Ref.~\cite{li2015atp}  to the time series $\hat{\alpha}(t)$ and $\hat{K}(t)$ obtained in Section~\ref{sec:alpha(t)}.
The authors of Ref.~\cite{li2015atp} improved the method proposed in Ref.~\cite{taylor2000change}, casting CP detection as a hypothesis testing problem and applied it to analyse the rotary motion of the F$_1$ molecular motor.
The algorithm requires setting a confidence level for accepting the hypothesis that there was a changepoint, as we discuss in~\ref{app:CP}. Setting the optimal confidence level requires balancing the trade-off between missing CPs (false negative) and wrongly detecting CPs when there was not one (false positive). The optimal choice depends on the data under consideration as we discuss in~\ref{app:CP} and  Fig.\ref{fig:tpfpfn}. For the benchmark dataset, we find a confidence level of $99.99\%$ returns the best performance in terms of JSC.
The CPDA applied to the temporal profile $\hat{\alpha}(t)$ returns consistent results, even though it displays a tendency to detect false positives CPs, as exemplified in Fig.\ref{fig:fig4}(a). 
The $\hat{K}(t)$ profile is less suitable for the CPDA as it accentuates the tendency to return false positives, as shown in Fig.\ref{fig:fig4}(b).
 We, therefore, focus on the performance of the CPDA applied to the  $\hat{\alpha}(t)$ profile. The CDPA does not reach the performance of \texttt{sequandi} and achieves a $\textrm{JSC}_\textrm{CP}=0.2470$ (roughly half of the one of \texttt{sequandi}) and $\textrm{RMSE}_\textrm{CP}=3.065$ (roughly twice the one of \texttt{sequandi}). 
However, despite its sensitivity to fluctuations in $\hat{\alpha}(t)$ and its tendency to return false positive detections, the CPDA outperforms \texttt{sequandi} for trajectories displaying many CPs, as we discuss in Section~\ref{sec:andi_results}.

\section{\texttt{sequandi} and the AnDi challenge 2024}\label{sec:andi_results}
In this Section, we describe how the \texttt{sequandi} network, which is proficient at capturing the time-dependent point-wise behaviour of the diffusive parameters, performs on the metrics evaluated by AnDi2024. 

\subsection{Structure of AnDi2024 data and task}
In AnDi2024, the goal is to infer the anomalous diffusion exponent $\alpha$ and generalised diffusion coefficient $K$ for each of the $N$ segments, characterised by different diffusive states that may be present in a single trajectory ($\alpha_i$ and $K_i$ for $i=0,...,N-1$). Also fundamental is the changepoint time $t_i^{\textrm{CP}}$ , \textit{i.e.}, the times at which the system switches between diffusive modes. 
All together, each trajectory is characterised by the vector containing the list of all the anomalous diffusion exponents, generalised diffusion coefficients and CPs (one per diffusive segment):
\begin{equation}
    [\alpha_0, K_0, t^{\textrm{CP}}_0, ..., \alpha_i, K_i,t^{\textrm{CP}}_i,...,\alpha_{N-1},K_{N-1},T-1]\,. \label{eq: submissionVec}
\end{equation}
In the final, so-called Challenge Phase of AnDi2024, the data was provided in folders of 12 different simulated experiments, each containing trajectories with unique $\alpha$, $K$ and CPs, drawn from a distribution specific to that experiment, which is chosen from one of the diffusion models described in Section~\ref{sec:intro}. 
\subsection{Performance metrics of AnDi2024}\label{sec: AnDiMetrics}
The pairing of the different diffusive segments is discussed in Section~\ref{sec:t_res}.
In addition to the JSC and the RMSE for CPs, AnDi2024 evaluates the performance of the inference of  $\alpha$ and $K$. 
This is evaluated for paired segments as 
\begin{eqnarray}
    \textrm{MAE}_{\alpha} = \frac{1}{N_{TP}}\sum_{\textrm{paired $i, j$}}^{N_{TP}} \left| \alpha_i-\hat \alpha_j \right|, \label{eq:mae_alpha} \\ \textrm{MSLE}_{K} = \frac{1}{N_{TP}}\sum_{\textrm{paired $i, j$}}^{N_{TP}}\left[ \log{\left[K_i+1\right]} - \log{[\hat K_j+1]} \right]^2\,. \label{eq:msle_K}
\end{eqnarray}
This is the error between the ground truths in Eq.~\eqref{eq: submissionVec} and segment-wise predictions $\hat\alpha_i, \hat K_i$, which we calculate in the next Section. We note that the calculation is over all of the identified TPs in the dataset together.

\subsection{Performances of \texttt{sequandi} on AnDi2024}\label{sec: segmentAK}
Our method, 
\texttt{sequandi}, is designed to return the time-dependent point-wise prediction $\hat\alpha(t),\hat K(t)$. 
We therefore need to process the output of \texttt{sequandi} to obtain the predictions required by AnDi2024, \textit{i.e.}, the vector of Eq.~\eqref{eq: submissionVec}.
To this aim, we first divide a trajectory in segments based on the $\hat t^{\textrm{CP}}_{i}$ predicted by the network described in Section~\ref{sec:cp}. 
We then compute the time average of  $\hat\alpha(t)$ and $\hat K(t)$, within each predicted segment $i$,
\begin{equation}
    \hat \alpha_{i} = \dfrac{\sum_{t=\hat t^{\textrm{CP}}_{i}}^{\hat t^{\textrm{CP}}_{i+1}-1} \hat\alpha(t)}{(\hat t^{\textrm{CP}}_{i+1}-1)-\hat t^{\textrm{CP}}_{i}}, \qquad \hat K_{i} = \dfrac{\sum_{t=\hat t^{\textrm{CP}}_{i}}^{\hat t^{\textrm{CP}}_{i+1}-1} \hat K(t)}{(\hat t^{\textrm{CP}}_{i+1}-1)-\hat t^{\textrm{CP}}_{i}}\,.
\label{eq:alphaK_seg}
\end{equation}
\begin{figure}[h!]
	\begin{center}
	\includegraphics[width=\textwidth]{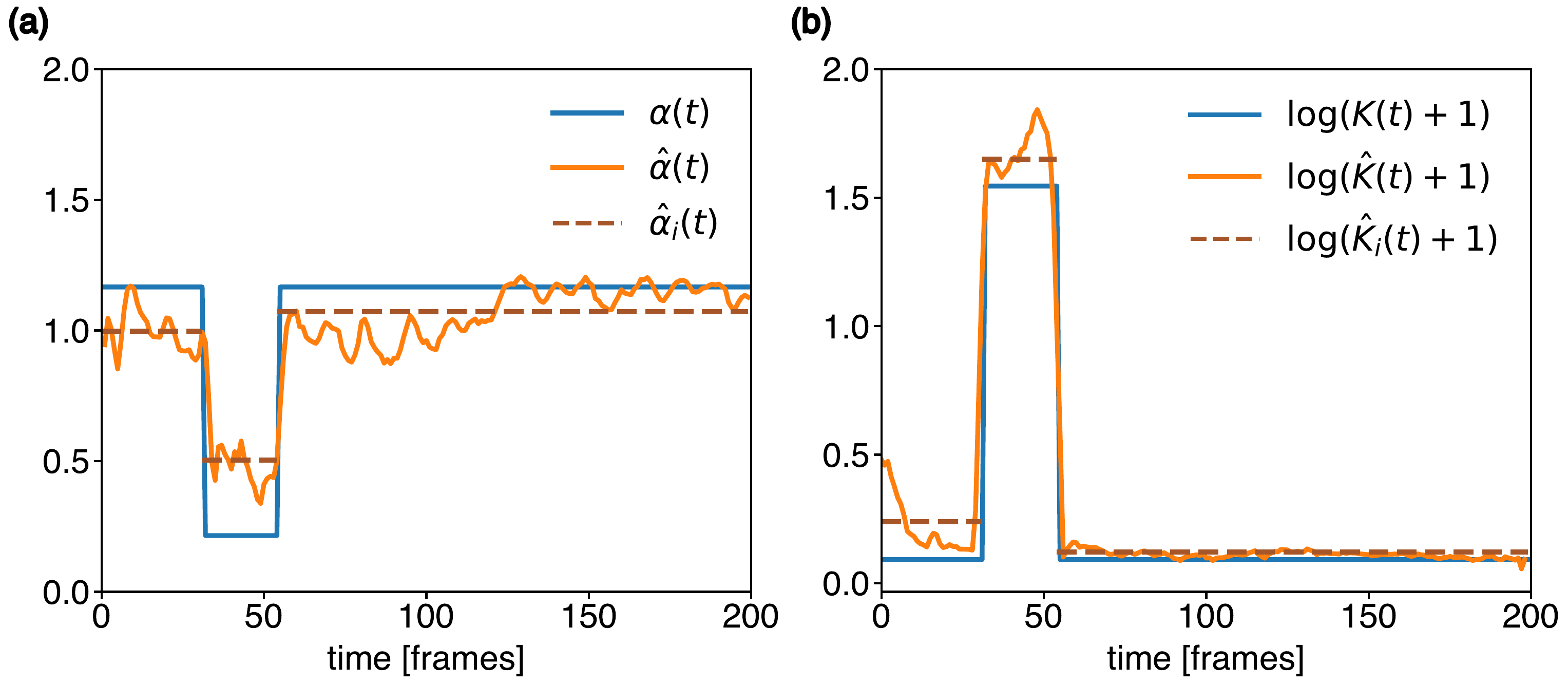}
	\end{center}
	\caption{{\bf Segmentation of $\hat\alpha(t)$ and $\hat K(t)$.}
        We segment the trajectory shown in Fig.~\ref{fig:fig1} based on the CPs predicted by \texttt{sequandi} $\hat t^{\textrm{CP}}$, which were indicated by the solid black lines shown in Fig.~\ref{fig:fig4}.
{\bf (a)} As in Fig.~\ref{fig:fig3}, the orange line is $\hat\alpha(t)$, and the blue piece-wise constant line is the ground truth $\alpha(t)$. The dashed brown line is the time averaged exponent for each segment $\hat \alpha_{i}$, following Eq.~\eqref{eq:alphaK_seg}.
    {\bf (b)} As in Fig.~\ref{fig:fig3}, the orange line is 
    $\log(\hat{K}(t)+1)$ and
    the blue piece-wise constant line is the ground truth $\log(K(t)+1)$.
    The dashed brown line is based on the time averaged generalised diffusion for each segment $\hat{K}_i$, following Eq.~\eqref{eq:alphaK_seg}, and is $\log(\hat{K}_i+1)$.
     }\label{fig:alphaK_i}
\end{figure}

Figure~\ref{fig:alphaK_i} shows the segment-wise predictions $\hat \alpha_{i}$ and $\hat K_{i}$ for the trajectory shown in Fig.~\ref{fig:fig1}, together with the point-wise predictions $\hat\alpha(t)$ and $\hat K(t)$ also shown in Fig.~\ref{fig:fig3}.
In this case, due to accurate predictions $\hat t^{\textrm{CP}}_{i}$ by \texttt{sequandi}, the $\hat\alpha_i, \hat K_i$ are not far from the ground truth. However, in general, inaccuracies in predicting the CPs
propagate and amplify, yielding inaccurate $\hat\alpha_i$ and $\hat K_i$ from  
what may otherwise have been informative and accurate $\hat \alpha(t)$ and $\hat K(t)$.\\

\begin{figure}[h]
	\begin{center}
	\includegraphics[width=\textwidth]{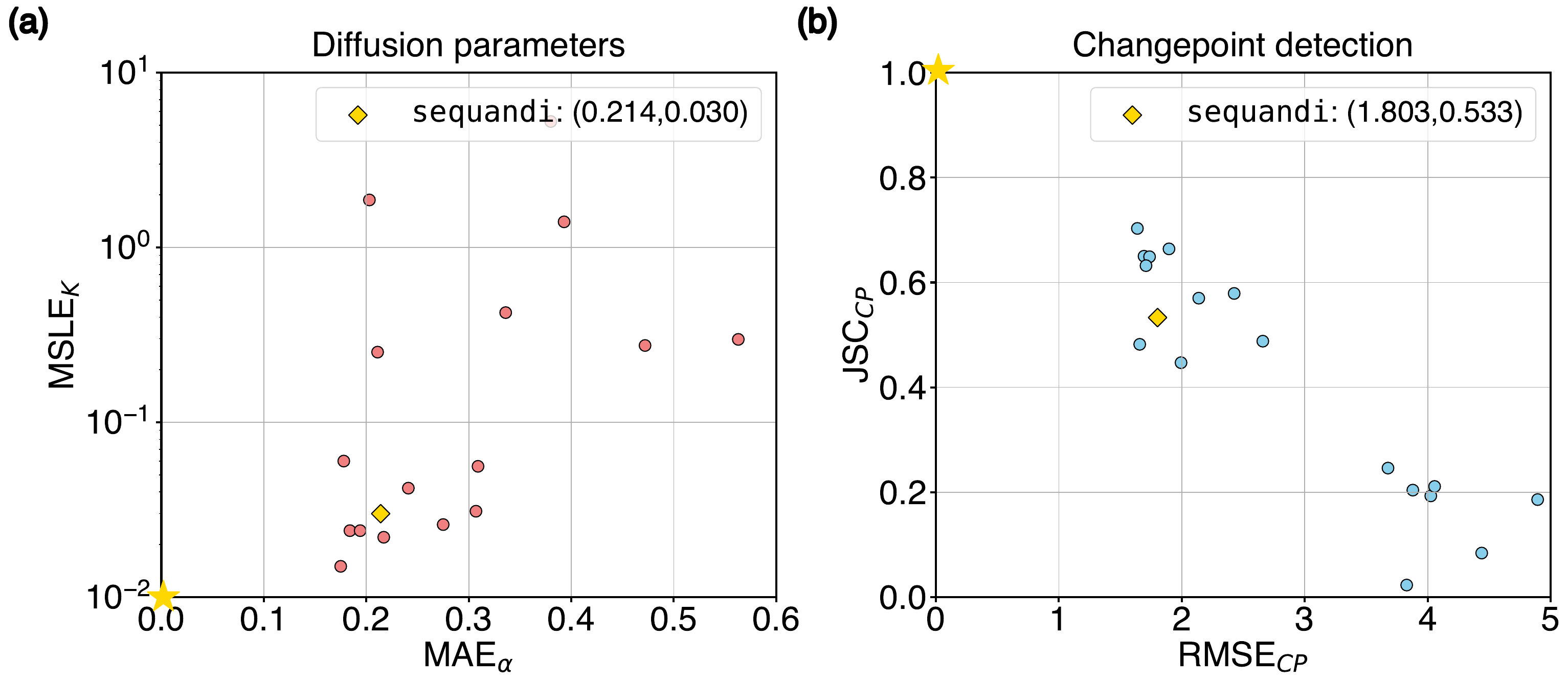}
	\end{center}
	\caption{{\bf Performance of \texttt{sequandi} in the 2024 AnDi Challenge.}  {\bf (a)} For the inference of the diffusion parameters, the MSLE$_K$ is plotted against the  MAE$_\alpha$. The star in the bottom left corner marks the performance of a hypothetical perfect prediction. The circles show the performances of the other methods participating in the 2024 AnDi challenge and the rhombus the one of \texttt{sequandi}.
    {\bf (b)} For the detection and location of CPs, the JSC is plotted against the  RMSE. The star in the top left corner marks the performance of a hypothetical perfect prediction. Again, the circles show the performances of the other methods participating in the challenge and the rhombus the one of \texttt{sequandi}.} 
	\label{fig:andi_score}
\end{figure}
In AnDi2024,
\texttt{sequandi}, under the KCL team name, finished $8^{\textrm{th}}$ overall in the Single-trajectory Task. 
To evaluate the performance in estimating the diffusion parameters, it is instructive to plot the $\textrm{MAE}_\alpha$ vs the $\textrm{MSLE}_K$ as shown in Fig.~\ref{fig:andi_score}(a).
In such a plot, a perfect prediction is located in the left bottom corner [marked by a star in Fig.~\ref{fig:andi_score}(a)].
The method of Ref.~\cite{asghar2025unet3anomalousdiffusion} outperforms the other ones, and \texttt{sequandi} is close to the cluster of runner-up methods.

For the changepoint detection and location, we plot the JSC vs the RMSE in Fig.~\ref{fig:andi_score}(b). The optimal prediction in this plot is on the top left corner [marked by a star in Fig.~\ref{fig:andi_score}(b)]. Also in this task, the best method is the one of Ref.~\cite{asghar2025unet3anomalousdiffusion} followed by a cluster of 4 runner-up methods. \texttt{sequandi} is located after the runner-up cluster.
\subsection{Benchmark dataset}\label{sec:bench}
To gain further insight into the capabilities of \texttt{sequandi}, we took advantage of the AnDi2024 benchmark dataset~\cite{munoz_gil_2024_benchmark}.
This dataset is similar to the 2024 AnDi Challenge one and contains 9 different simulated experiments:
\begin{itemize}
    \item \textit{Experiment 1:} A 3-state model inspired by the experiments of~\cite{Sungkaworn17}. All states are characterised by $\alpha$ close to the normal diffusion $\alpha\sim1$ but each state has a different $K$, see the $x$ axis of Fig.~\ref{fig:app_alpha} and~\ref{fig:app_K}.
    Switching between states occurs rapidly.
    \item \textit{Experiments 2 and 5:} Two dimerisation models. Experiment 2 is inspired by~\cite{Low-Nam11}, where two particles can bind and unbind, therefore being present in a bound and unbound state of similar $\alpha\sim1$ but different $K$. In Experiment 5, one of the states has $\alpha\sim1$, and the other one is strongly sub-diffusive $\alpha<0.25$.
    \item \textit{Experiments 3 and 9:} Two trapping models characterised by a state where there is no motion ($\alpha\sim0$, $K\sim0$).  Outside of the immobile state, Experiment 3 features diffusion close to normal $\alpha\sim1$. In Experiment 9, motion is nearly ballistic, with transitions out of the trapped state occurring much more frequently.
    \item \textit{Experiment 4:} Confinement model with distinct, slow confined state. 
    \item \textit{Experiments 6 and 7:} Dimerisation and 2-state model, respectively. They share identical diffusive parameters, but with different underlying experimental constraints.
    \item \textit{Experiment 8:} Single-state diffusion with broad distributions of $\alpha$ and $K$.
\end{itemize}
The distribution of the various diffusive parameters can be seen by observing the ground truth values of the plots in Figs.~\ref{fig:app_alpha} and~\ref{fig:app_K}.\\

\begin{figure}[h! t]
	\begin{center}
	\includegraphics[width=\textwidth]{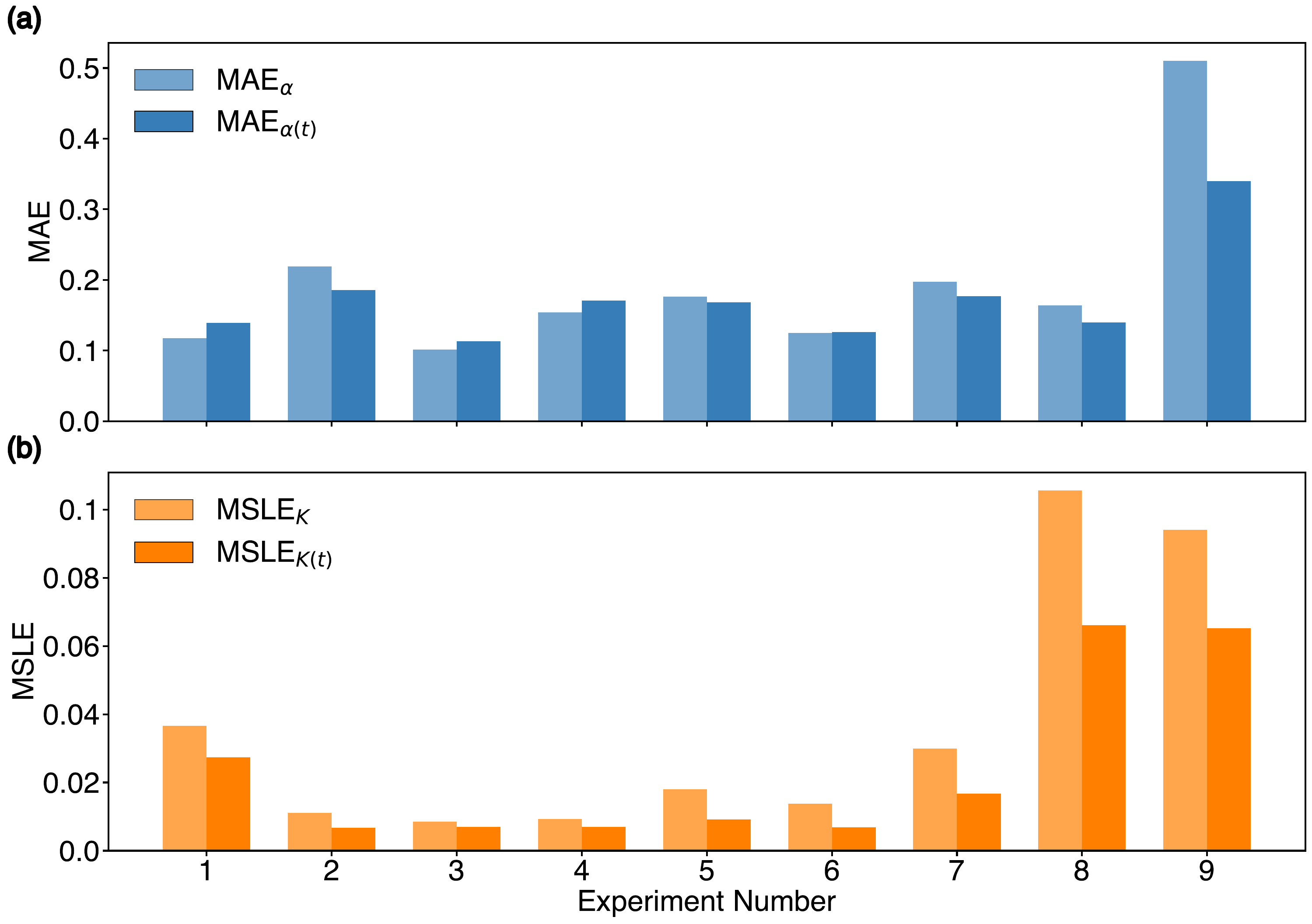}
	\end{center}
	\caption{{\bf Point-wise (\texttt{sequandi}) and segment-wise performances for diffusion inference on each experiment of the benchmark dataset.} 
	{\bf (a)} Mean absolute error of $\alpha$ per experiment for the segment-wise predictions $\hat{\alpha}_i$, $\textrm{MAE}_{\alpha}$, and for the point-wise sequential predictions $\textrm{MAE}_{\alpha (t)}$. {\bf (b)} Mean squared logarithmic error (MSLE) of $K$, per experiment for the segment-wise predictions $\hat{K}_i$, $\textrm{MSLE}_{K}$, and for the point-wise sequential predictions $\textrm{MSLE}_{K (t)}$.} 
	\label{fig:exp_aK}
\end{figure}

As stated before, \texttt{sequandi} is designed to perform a time-dependent, point-wise inference of $\hat{\alpha}(t)$ and $\hat{K}(t)$. This task is, in principle, more difficult than estimating just the diffusive parameters within a segment of a trajectory $\alpha_i$ and $K_i$. 
The point-wise metrics $\textrm{MAE}_{\alpha(t)}, \textrm{MSLE}_{K(t)}$ are more stringent than the segment-wise  $\textrm{MAE}_{\alpha}, \textrm{MSLE}_{K}$. 
However, processing the point-wise output of \texttt{sequandi}, $\hat{\alpha}(t)$ and $\hat{K}(t)$, to obtain the segment-wise $\hat{\alpha}_i$ and $\hat{K}_i$ is prone to errors due to the uncertainty in estimating the CPs.
As a result, we find that, on the benchmark data, the segment-wise metrics are $\textrm{MAE}_\alpha=0.206, \textrm{MSLE}_K=0.027$\footnote{Throughout the paper, the metrics on the benchmark dataset are averages of the metrics obtained on each of the 9 experiments, weighted by the number of trajectories belonging to each experiment. The AnDi2024 segment-wise metrics expressed in Eq.~\eqref{eq:mae_alpha} and~\eqref{eq:msle_K} calculate the error across all of the paired segments in the data together. To make closer contact with this metrics, in this Section, instead of first taking $\textrm{MAE}_{\alpha(t)}, \textrm{MSLE}_{K(t)}$ at the trajectory level as in Eq.~\eqref{eq:mae_alphat} and~\eqref{eq:msle_Kt}, we perform the calculation over all of the flattened times $t$ across all the trajectories.}, which are significantly worse than the point-wise metrics from the set of flattened $\hat \alpha(t)$ and $\hat K(t)$ sequences ($\textrm{MAE}_{\alpha(t)}=0.176, \textrm{MSLE}_{K(t)}=0.014$).
 
\begin{figure}[h! t]
	\begin{center}
	\includegraphics[width=\textwidth]{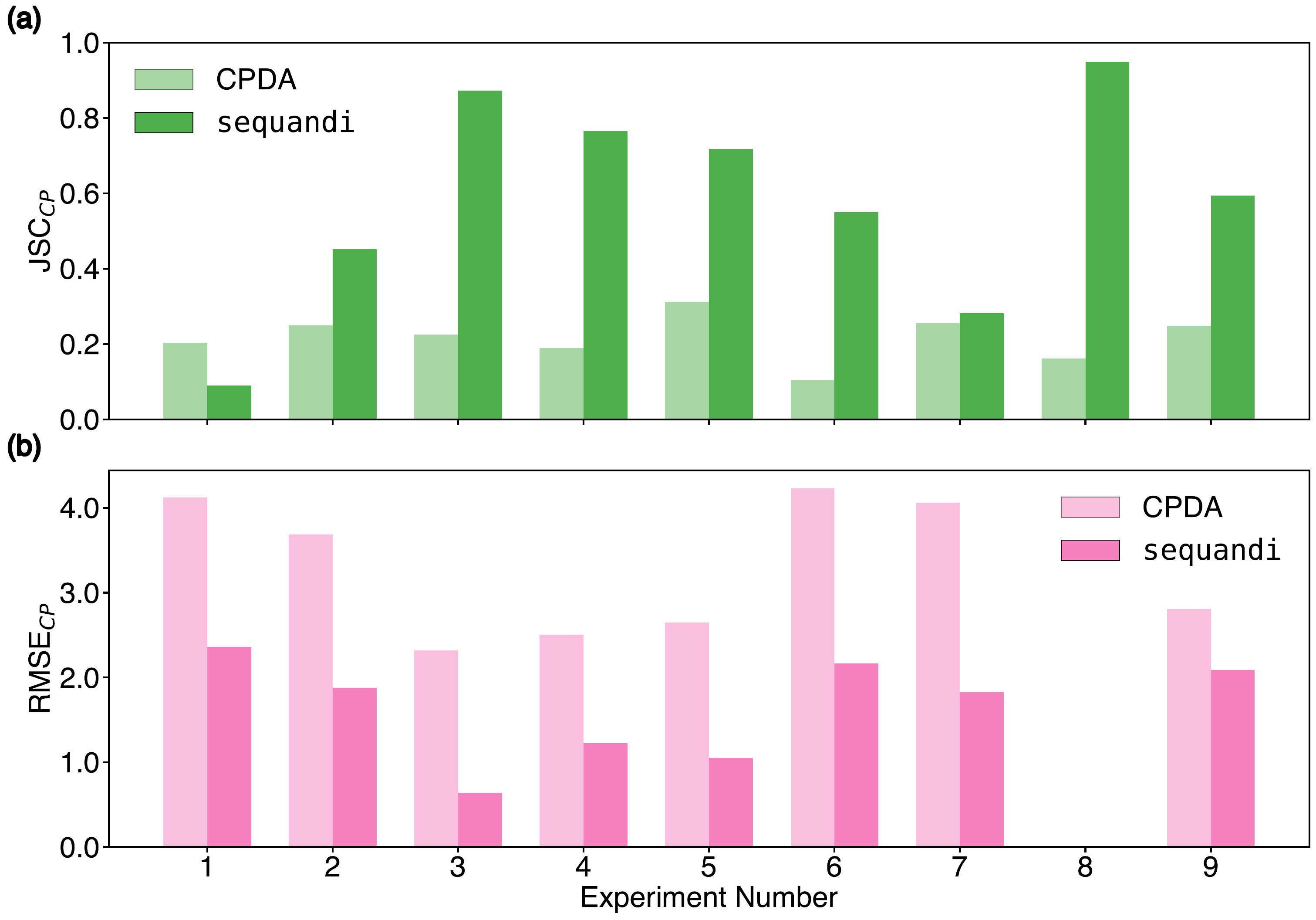}
	\end{center}
	\caption{{\bf \texttt{sequandi} and CPDA performances on CP detection and location on each experiment of the benchmark dataset} 
	{\bf (a)} Comparison of Jaccard Similarity Coefficient of changepoint detections $\textrm{JSC}_\textrm{CP}$, between the CPDA with confidence level 99.99, and by the $\texttt{sequandi}$ neural network. $0\le \textrm{JSC}\le 1$, with 1 being perfect detection. {\bf (b)} Root Mean Squared Error of true positive changepoint detected $\textrm{RMSE}_\textrm{CP}$, by $\texttt{sequandi}$ and the CPDA.} 
	\label{fig:exp_CP}
\end{figure}

 It is interesting to analyse each experiment separately, as reported in Fig.~\ref{fig:exp_aK}.
Figure~\ref{fig:exp_aK}(b) shows that the generalised diffusion coefficient $K$ is more accurate for the point-wise estimate than for the segment-wise one. 
 The picture is more nuanced for the inference of the anomalous diffusion exponent $\alpha$, shown in Fig.~\ref{fig:exp_aK}(a). In Experiments 3 and 4, where the CP detection is accurate (see Fig.~\ref{fig:exp_CP}), the segment-wise prediction is more accurate than the point-wise one. Experiment 1 is exceptional because it displays many challenging CP detections, which lead to poor performance by \texttt{sequandi}. In this experiment, the segment-wise prediction outperforms the point-wise one, perhaps because the segments had a very similar ground truth $\alpha\sim1$. 

 Overall, for the inference of $\alpha$ and $K$, \texttt{sequandi} performs well and evenly throughout different types of experiments with two exceptions. Experiment 9 is the most challenging one because of its unique combination of immobile states and nearly ballistic ones that are edge cases in the training set, see Fig.~\ref{fig:app_training}.
 Experiment 8 is also challenging for the $K$ inference because it displays a single state but covers a vast range of $K$, which was not optimised during training (Compare Fig.~\ref{fig:app_training} with Fig.~\ref{fig:app_K}).
In Figs.~\ref{fig:app_alpha} and~\ref{fig:app_K}, we report how individual $\hat \alpha_i$ and $\hat K_i$ predictions compare to their ground truths on an experiment-wise basis.

We now turn our attention to the CP detection and location and the comparison between the predictions of \texttt{sequandi} and those of the CPDA, which we report in Fig.~\ref{fig:exp_CP}.
As anticipated before, the most challenging CP detections are in Experiment 1, which features frequent transitions between 3 diffusive states, followed by Experiments 2, and 7.
\texttt{sequandi} systematically outperforms the CPDA applied to $\hat{\alpha}(t)$ except in Experiment 1.
The sensitive nature of the CPDA is confirmed by its ability to detect more changepoints than \texttt{sequandi} in Experiment 1 and by grossly underperforming in Experiment 8, where there are no CP. We report a detailed analysis of the impact of the confidence levels on the performance of the CPDA in Fig.~\ref{fig:app_tCP}.

\section{Conclusion and discussion}
This article presents \texttt{sequandi}, a method based on LSTM recurrent neural networks to study single trajectories undergoing anomalous diffusion and switching between different diffusive states. Leveraging the properties of RNNs, \texttt{sequandi} can analyse trajectories of varying length and returns a point-wise time-dependent estimate of the diffusive properties of the trajectory in terms of its anomalous diffusion exponent $\alpha$ and generalised diffusion coefficient $K$, $\hat{\alpha}(t)$ and $\hat{K}(t)$, respectively.
Such a time-dependent estimate is well-suited to study trajectories that irregularly switch between different diffusive states. We complement this analysis by training another network to detect and locate the changepoints when a switch between different diffusive states occurs $\hat{t}^\textrm{CP}$. 
We compare the performance of this neural network to that of the Change Point Detection Algorithm (CPDA) proposed in Ref.~\cite{li2015atp} applied to the time-dependent estimate $\hat{\alpha}(t)$.
At variance with the network, the CPDA requires carefully setting a confidence level, which is a non-trivial step, whose optimal solution depends on the task at hand. In general, \texttt{sequandi} outperforms the CPDA applied to $\hat{\alpha}(t)$.
A possible way to improve the CPDA performance is to apply it to a combination of $\hat{\alpha}(t)$ and $\hat{K}(t)$ or to use also the uncertainty uncertainty $\sigma^2_\alpha(t)$, which one would have to compute, as done in Ref.~\cite{seckler2022bayesian,seckler2024change}.
Another possibility would be pooling the CP detections of  \texttt{sequandi} and the CPDA. This would require careful tuning, but it has the potential to mitigate the opposite tendencies towards false positives and false negatives the two approaches show in Fig.~\ref{fig:exp_CP}.
\\
We benchmark the performances of \texttt{sequandi} on the 2024 Anomalous Diffusion Challenge dataset, which is based on detecting the changepoints and predicting the $\alpha$ and $K$ for each segment between changepoints. This requires combining the results of the network that detects the changepoints $\hat{t}_\textrm{CP}$ with the one that infers $\hat{\alpha}(t)$, and $\hat{K}(t)$ and then taking an average over the identified segments. This procedure is not error-free and degrades the performances of the method. However, \texttt{sequandi} was significantly outperformed only by the method of Ref.~\cite{asghar2025unet3anomalousdiffusion} and performed comparably to the runner-up methods (see Fig.~\ref{fig:andi_score}).\\
Our study underscores the inherent challenge in integrating algorithmic and neural network-based approaches without propagating errors. This highlights that, when possible, performing the whole analysis with a neural network is advantageous. This is because the network automatically learns ways of compensating uncertainties and errors to return optimal predictions, as shown, for instance, in Ref.~\cite{Argun_2021}.\\ 
Nonetheless, we stress that the point-wise inferences $\hat{\alpha}(t)$ and $\hat{K}(t)$ of \texttt{sequandi} are accurate across a broad scope of systems with switching diffusive behaviour.\\ 
Since some experimental data comes from three-dimensional tracking, it may be interesting to extend the method to address three-dimensional trajectories. This extension would simply require adapting the data pre-processing and training the networks on three-dimensional data.
It would also be interesting to explore the performance of different neural network architectures.
Among these, transformers~\cite{vaswani2017attention} may be especially promising, given their success in sequence modelling tasks, including natural language processing and time series forecasting. Their application to the study of anomalous diffusion has not been straightforward~\cite{qu2024semantic} due to difficulties in encoding the input trajectories~\cite{PhysRevE.107.034138}, which have been mitigated by hybrid approaches, including initial CNN layers~\cite{requena2023inferring, firbas2023characterization, feng2025enhancing}.
The AnDi2024 database will provide an objective testbed to compare the capabilities of different approaches, and has already highlighted the potential of CNN~\cite{asghar2025unet3anomalousdiffusion}.

\subsection*{Acknowledgements}
SB wishes to thank Chun-Biu Li for discussions on the changepoint algorithm of Ref.~\cite{li2015atp}.
XQ was financially supported by the China Scholarship Council through the King’s-China Scholarship Council PhD Scholarship programme (K-CSC). 
This work benefited from access to the Computational Research, Engineering and Technology Environment (CREATE) at King’s College London.
We are grateful to the UK Materials and Molecular Modelling Hub for computational resources, which is partially funded by EPSRC (EP/T022213/1, EP/W032260/1 and EP/P020194/1. We acknowledge the support of Alejandro Santana Bonilla.
\appendix

\section{Training and Testing Procedures}\label{app:traintest}
In this Appendix, we provide additional details about the structure of the $2.5\times 10^6$ trajectories that we used for training \texttt{sequandi}.
and, analysing an additional, independent test dataset, we show that the network is correctly trained (not overfitting).
\begin{figure}[h]
	\begin{center}
	\includegraphics[width=0.405\textwidth]{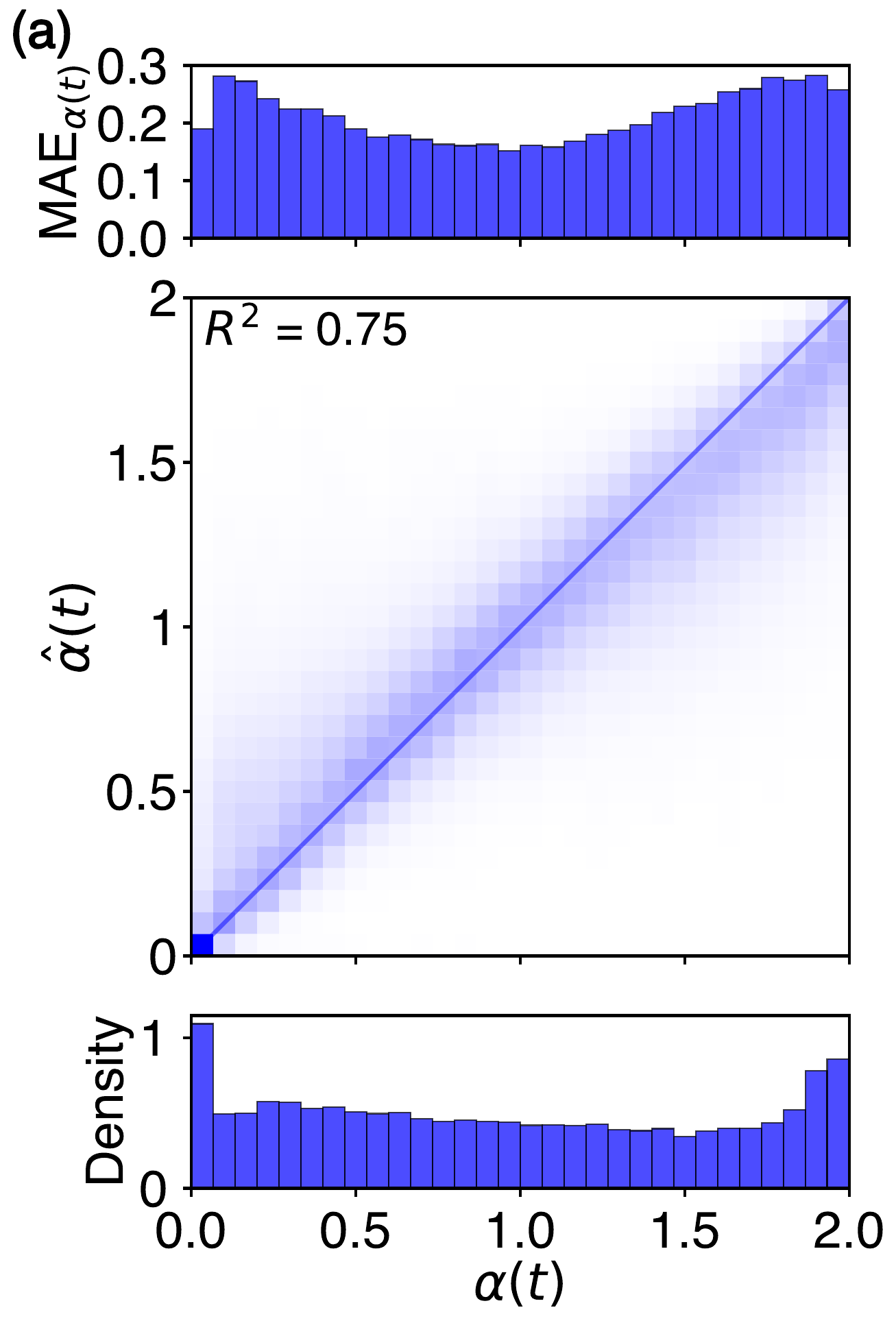}
        \includegraphics[width=0.4\textwidth]{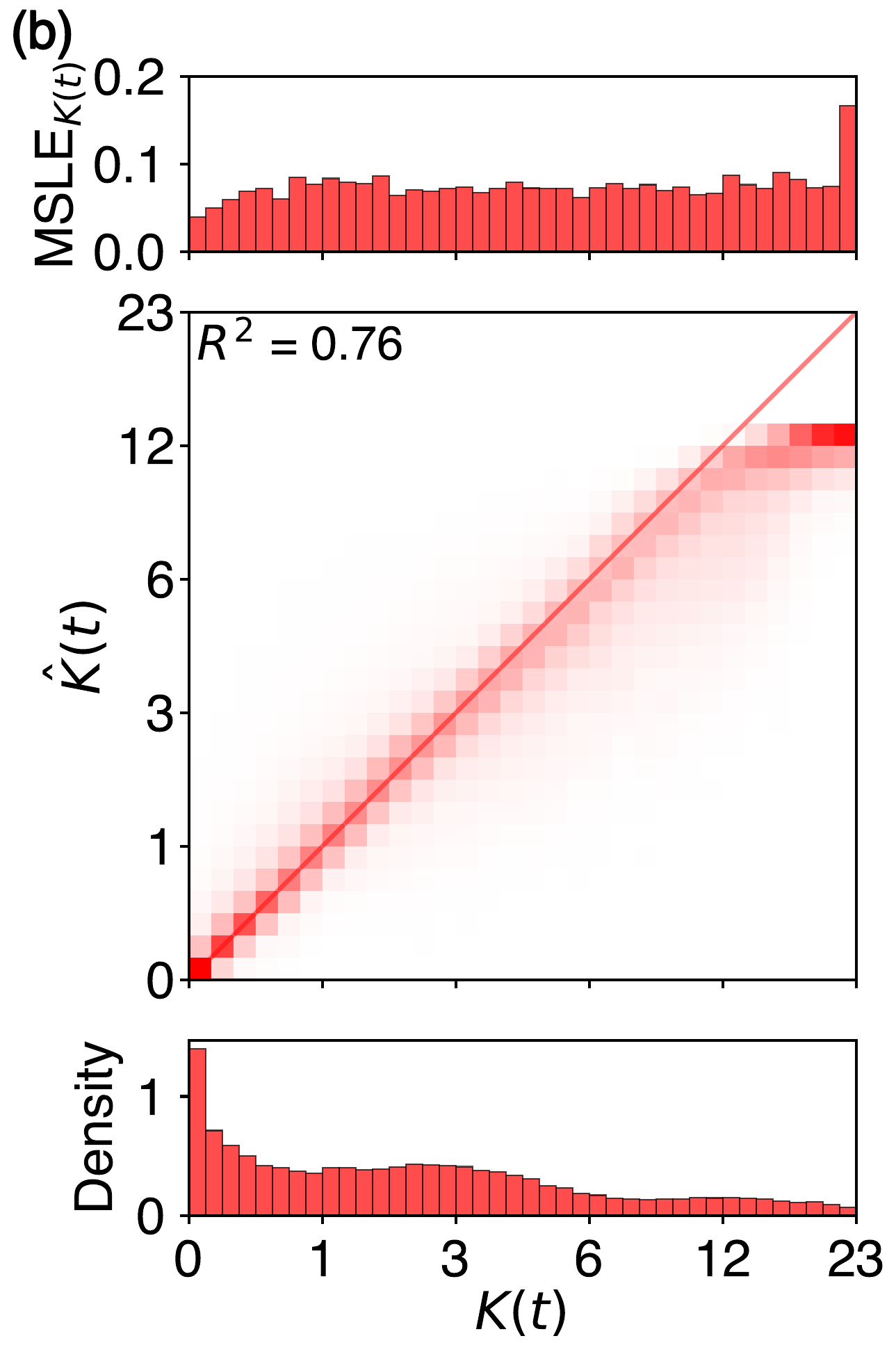}
	\end{center}
	\caption{{\bf Training dataset distributions and predictions.}
    \textit{Bottom row histograms}: distribution of the ground truth \textit{point-wise} parameters of \texttt{sequandi}'s training dataset, {\bf(a)} anomalous diffusion exponent $\alpha (t)$ and {\bf(b)} generalised diffusion coefficient $K(t)$. Note the logarithmic scaling of $K$.
    \textit{Density maps}: Predicted $\hat{\alpha}(t)$ and $\hat{K}(t)$ vs their respective ground truths. Colours are normalised in each column with respect to the ground truth probabilities below. \textit{Top row histograms}: average mean absolute error (MAE) and mean squared logarithmic error (MSLE), of the points in each ground truth column. Per trajectory, the error averages over the entire dataset are $\textrm{MAE}_{\alpha(t)} = 0.247$ and $\textrm{MSLE}_{K(t)}=0.140$.}
	\label{fig:app_training}
\end{figure}
To generate the training set, we first uniformly sample from the five experiments in the \texttt{andi-datasets} Python library in $\alpha$ and $\log{K}$. As mentioned in the main text, in supervised regression tasks, the values at the edges of the considered intervals are often difficult to estimate and an ``S-shape'' often occurs, where the lowest values are overestimated and the highest ones underestimated~\cite{Argun_2021}. To mitigate this issue and to have more training data for the most challenging cases, which where frequent in the Andi2024 dataset, we add trajectories with $\alpha$ close to 2 and small $K$. The parameter distributions are shown in Fig.~\ref{fig:app_training} together with the corresponding point-wise
predictions. From the distributions in Fig.~\ref{fig:app_training}(a), we see the network excels for the central part of the $\alpha$ interval but underestimates the near ballistic case with $\alpha\approx2$. This is confirmed by the $\alpha$ dependence of the $\textrm{MAE}_{\alpha (t)}$, (the upper panel of Fig.~\ref{fig:app_training}(a)), which shows how the network performs best in the middle of the $\alpha$ range.
For the generalised diffusion coefficient $K$, the network trained well for  $K\lesssim 10$ but systematically underestimates for $K>10$, as shown in
 Fig.~\ref{fig:app_training}(b). This is likely because, despite sampling evenly the number of segments with $K_i$ in this regime, trajectories with high $K$ are shorter, because they leave the simulation's field of view more quickly. As a consequence, the piece-wise trajectories $K(t)$ are shorter, making them under-represented in the training dataset.\\
 We also report the
 $R^2$ metric, which evaluates how well the predictions fit a regression model given by the optimal diagonal. Precisely, for ground truth $y_i$ with mean $\bar y$,  and predictions $\hat y_i$
\begin{equation}
    R^2 = 1 - \frac{\Sigma_i(y_i-\hat y_i)^2}{\Sigma_i(y_i-\bar y)^2}
\end{equation}
This value ranges from 0 to 1, with 1 being the perfect fit. $\alpha$ and $K$ have a similar score with $R^2_{\alpha(t)}=0.75$ and $R^2_{K(t)}=0.76$.

\begin{figure}[h]
	\begin{center}
	\includegraphics[width=0.7\textwidth]{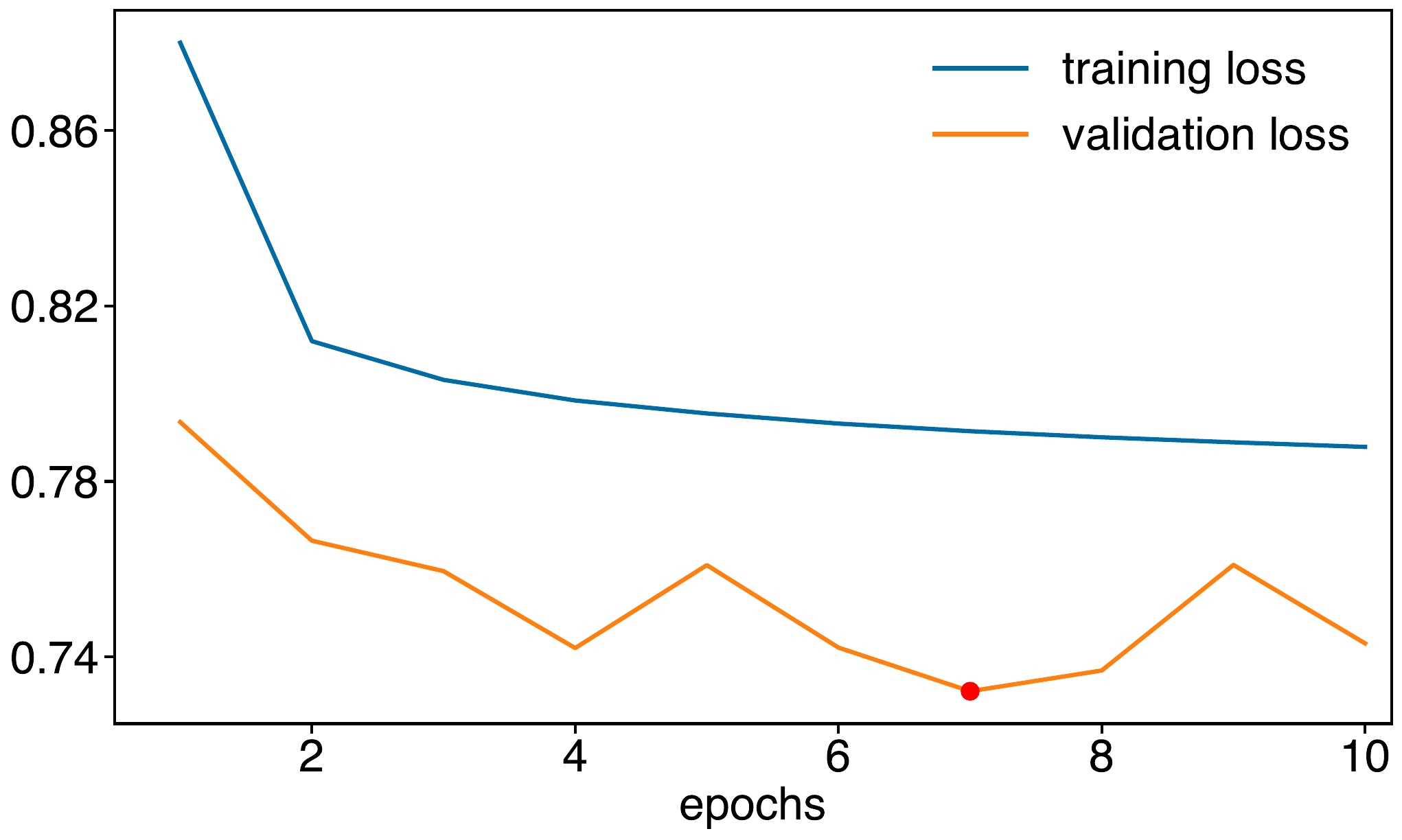}
	\end{center}
	\caption{{\bf Training and validation loss of the network.}
    Training and validation losses are shown for each epoch in the training procedure of the network that infers $\hat\alpha (t)$ and $\hat K(t)$. The red circle indicates the epoch the model reverts to when the early stopping is activated. Training loss is scaled by a third for illustration.} 
	\label{fig:app_losscurves}
\end{figure}

In Fig.~\ref{fig:app_losscurves}, we plot the training and validation losses as a function of the training epochs, for the network trained on the diffusion parameters. We implement early stopping to prevent overfitting, with a patience of 3 epochs, such that the training stopped in the 7th epoch after no improvements in this time-frame were seen. Training is performed on an HPC system using Tensorflow with CUDA, on an NVIDIA A100-SXM4-40GB GPU, and lasts 164 minutes per epoch on average.
\begin{figure}[h]
	\begin{center}
	\includegraphics[width=0.38\textwidth]{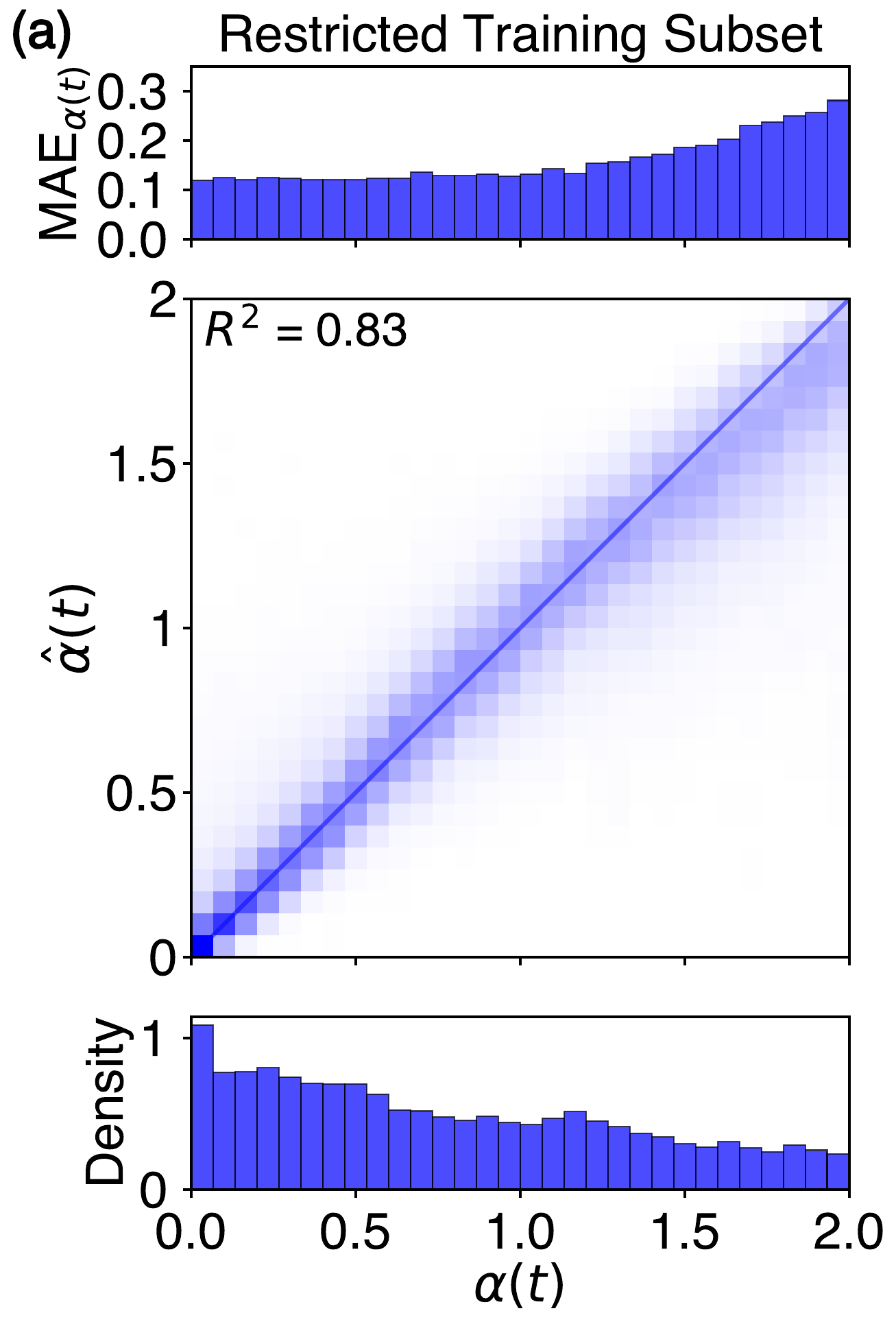}
        \includegraphics[width=0.38\textwidth]{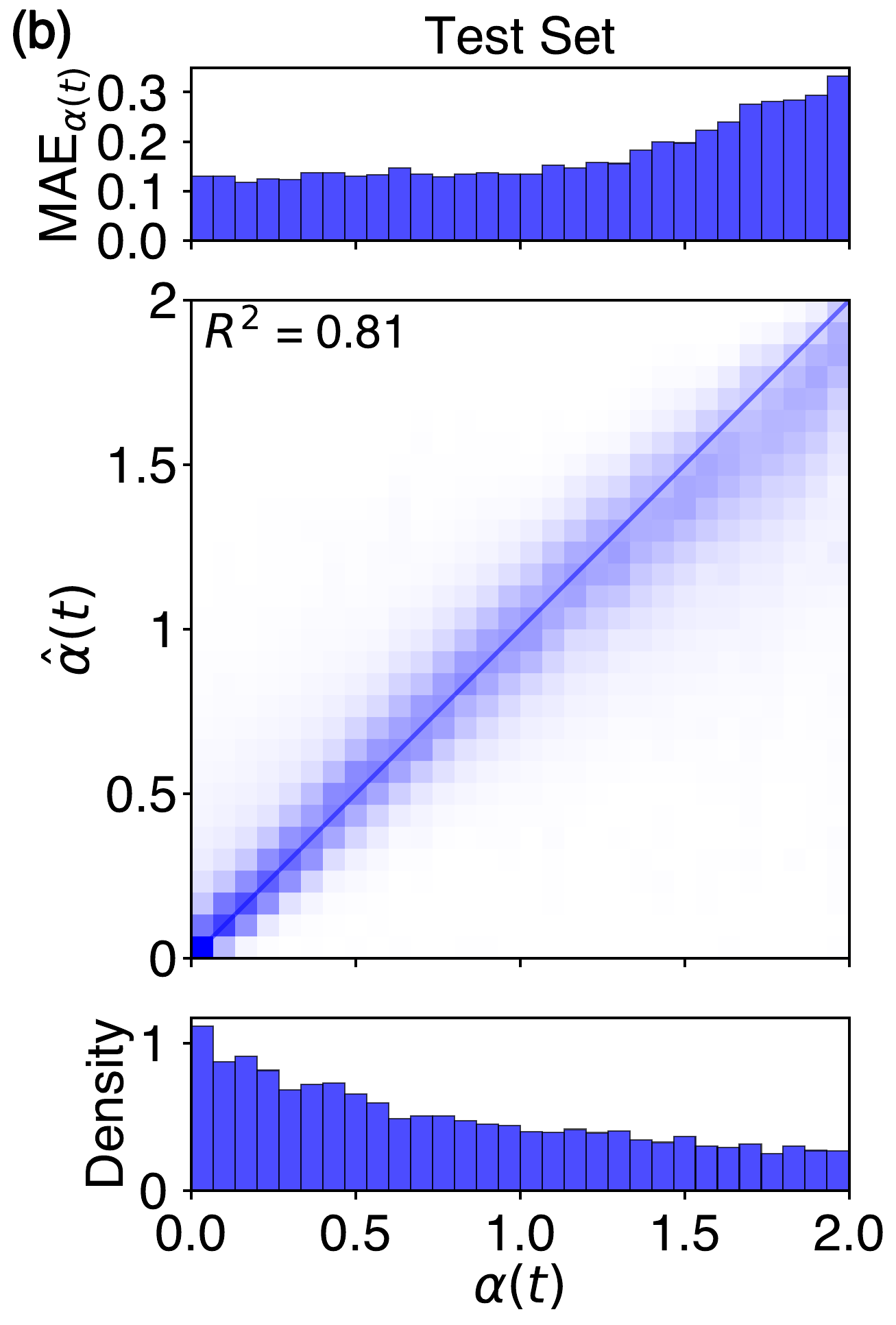}
        \includegraphics[width=0.385\textwidth]{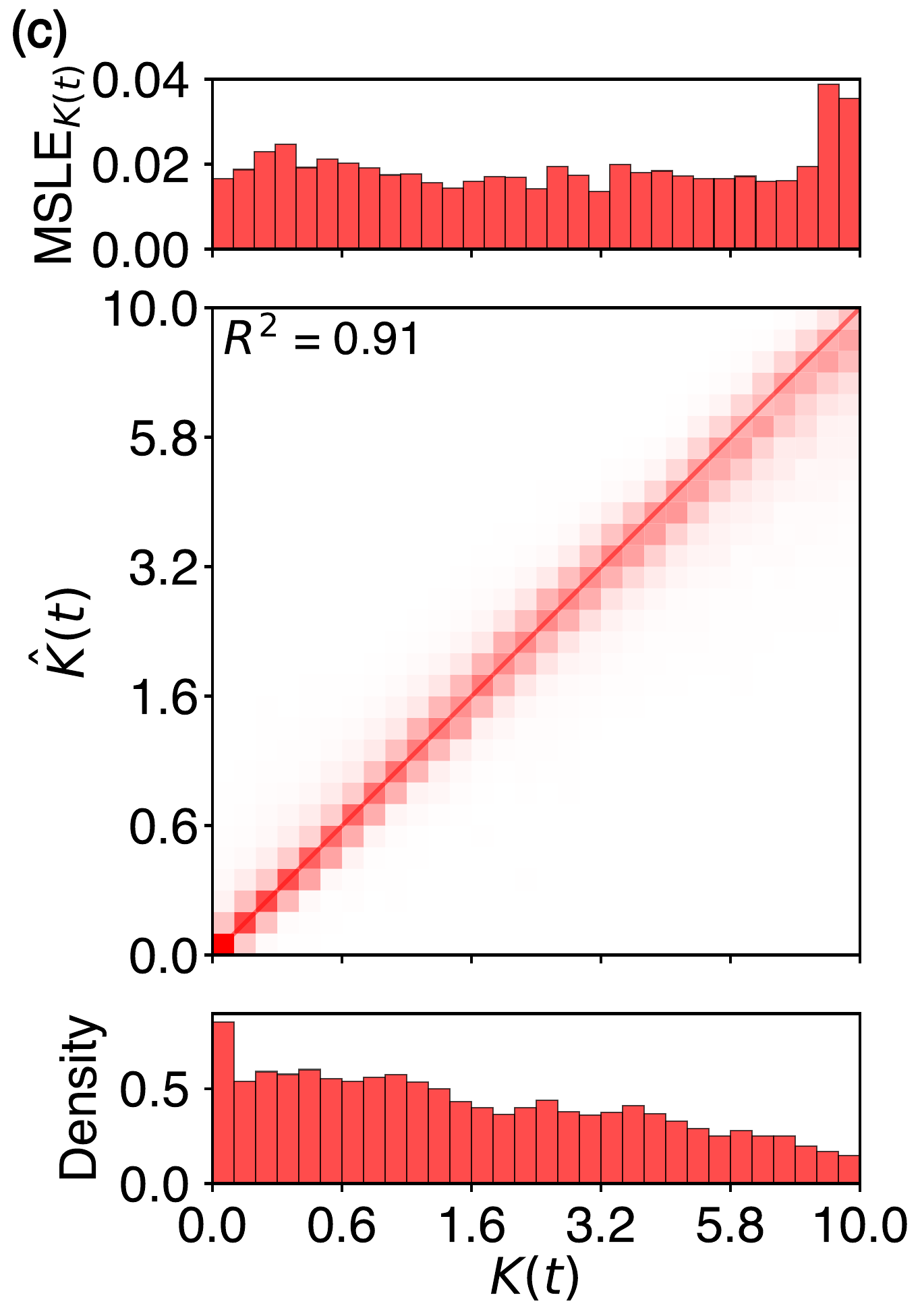}
        \includegraphics[width=0.385\textwidth]{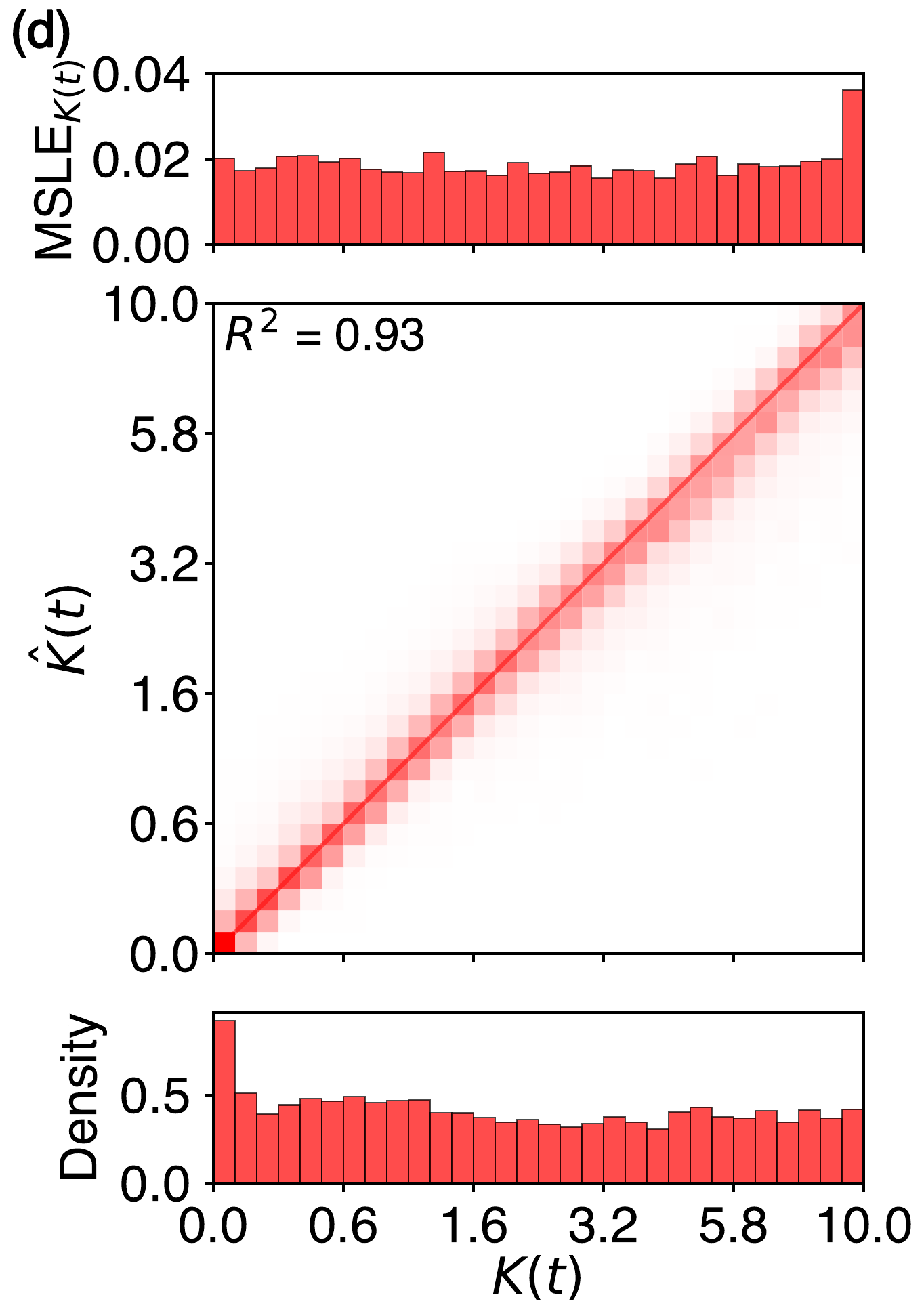}
	\end{center}
	\caption{{\bf Distributions and predictions of the Restricted training subset and the Test set.}
    Predictions vs ground-truth for $\alpha(t)$ [{\bf(a)} for the restricted training dataset, {\bf(b)} for the test dataset], and for $K(t)$ [{\bf(c)} for the restricted training dataset, {\bf(d)} for the test dataset].
    The top panels report histograms of the MAE and MSLE as a function of the ground truth. 
    The density maps report the predicted $\hat{\alpha}(t)$ and $\hat{K}(t)$ vs their respective ground truths. The density is normalised so that each column sums to one.
      The bottom panels histograms are the distributions of the ground truth parameters of the datasets. 
      Note the logarithmic scaling of $K$.
      } 
    \label{fig:app_restricted}
\end{figure}
To investigate the network's learning, we create an independent test dataset, on a balanced set, featuring an approximately uniform distribution of $\alpha$ with $0<\alpha<2$ and an approximately uniform distribution of $\log K$ for each segment.
Trajectories with large $\alpha$ and $K$ rapidly lead the field of view, so that they tend to be observed for a shorter time, leading to the decreasing distribution of $\alpha(t)$ and $K(t)$ shown in the lower panels of Fig.~\ref{fig:app_restricted}.
To assess overfitting, we compare the performance to a \textit{restricted subset} of the training set, with a comparable distribution of $\alpha$ and $K$.
Both the restricted training and test datasets have $2\times10^4$ trajectories, with the distributions of point-wise parameters illustrated in the bottom histograms of each subplot in Fig.~\ref{fig:app_restricted}. Note that, despite uniformly sampling the segment parameter distributions, there are some discrepancies between restricted and test sets in the point-wise distributions due to stochastic variations in the bulk experiments the trajectories are created within.
Comparing the restricted training and the test data set (left and right density maps in Fig.~\ref{fig:app_restricted}, respectively), we observe comparable performances, which are confirmed by the
 $\textrm{MAE}_{\alpha (t)}$ and $\textrm{MSLE}_{K(t)}$ metrics reported in Table~\ref{tab: 1}.
 \begin{table}[ht]
\centering
\begin{tabular}{|c|c|c|c|c|c|c|}
\hline
Dataset & $\textrm{MAE}_{\alpha(t)}$ & $\textrm{MSLE}_{K(t)}$ & $\textrm{RMSE}_{CP}$ & $\textrm{JSC}_{CP}$ & $\textrm{MAE}_{\alpha}$ & $\textrm{MSLE}_{K}$ \\
\hline
Training & 0.247 & 0.140 & 1.873 & 0.342 & 0.274 & 0.208 \\
Restricted Train & 0.185 & 0.045 & 1.455 & 0.705 & 0.180 & 0.061 \\
Test & 0.191 & 0.046 & 1.412 & 0.737 & 0.193 & 0.057 \\
Benchmark & 0.191 & 0.021 & 1.504 & 0.577 & 0.206 & 0.027 \\
\hline
\end{tabular}
\caption{Performance of \texttt{sequandi} based on metrics of predictions in different datasets. $\textrm{MAE}_{\alpha(t)}$ and $\textrm{MSLE}_{K(t)}$ are the average of point-wise mean absolute error in $\alpha$ and mean squared logarithmic error in $K$, respectively, of all trajectories in each dataset. $\textrm{RMSE}_{CP}$ is the root mean squared error between true and predicted changepoint times that are true positives (TPs), given in Eq.~\eqref{eq:RMSE}. $\textrm{JSC}_{CP}$ is the rate of these TPs (higher is better), given in Eq.~\eqref{eq:JSC}.  $\textrm{MAE}_{\alpha}$ and  $\textrm{MSLE}_{K}$ are the average errors of $\alpha$ and $K$ only in the ensemble of TP segment detections. }\label{tab: 1}
\end{table}
 These metrics (closely related to the ones used during training) show how the restricted training and the test datasets have similar performances, demonstrating that the training did not lead to overfitting. For completeness, we also report in Table~\ref{tab: 1} the metrics used in the AnDi challenge segment-wise inference, using the paired segments as described in Sections~\ref{sec:t_res} and ~\ref{sec: AnDiMetrics}. These evaluate predictions from the post-processed pooling of $\hat\alpha (t), \hat K(t)$ with the $\hat{t}^{\textrm{CP}}$ trained on a different network. These metrics are interesting for evaluating how the networks perform on different datasets but do not provide immediate information about overfitting.
\subsection{Computation times of predictions} 
 While training is a time-consuming process, using the trained networks to make predictions is relatively fast.
 To infer $\hat\alpha (t)$ and $\hat K(t)$, we must use the networks three times per trajectory set, as described in Section~\ref{sec: postproc}; once for the original trajectory, and twice more for a shift of one and two increments. The three predictions (with three time-step resolution) are pooled to obtain the refined $\hat\alpha (t)$ and $\hat K(t)$ traces with a single time-step resolution. Using the \texttt{sequandi} networks on a laptop (13th Gen Intel Core i7 processor, 32 GB RAM), it takes 8 minutes for the test dataset predictions pictured in Fig.~\ref{fig:app_restricted}(b) to be made. Comparatively, on this dataset it takes 20 minutes to find all $\hat\alpha (t)$ and $\hat K(t)$ using the sliding window TA-MSD technique with linear fits.


\section{Hyperparameter fine-tuning}\label{app:hyper}
We chose our architecture combining information from the literature,~\cite{argun2021simulation} and trial and error.
In this section, we compare our performance with the one of a network where we
 systematically perform a hyperparameter search with the hyperband method \cite{Li2018hyperband} in Keras. This is a tuning method which searches randomly at first, trying different networks for short amounts of time, before going deeper in the training with the best candidates. The candidates are whittled down successively like in a sports bracket, until the best one remains. We perform the search both on the network designed for diffusion parameter inference and for the one for changepoint locations' prediction, with 90 combinations of hyperparameters being tested in each case. The best architecture for the $\alpha (t)$ and $K(t)$ inference is: [LSTM Dimensions 320 and 160, recurrent dropout 0.2, dropout 0.1, learning rate 0.003]. For the $t^{\textrm{CP}}$ inference: [LSTM Dimensions 380 and 150, recurrent dropout 0.1, dropout 0.2, learning rate 0.0005]. These architectures are not too different from the one of \texttt{sequandi}. We train them on the full training set, and find that their performance is comparable to that of \texttt{sequandi}, as reported in Table~\ref{tab: 2} on the four datasets. Interestingly, \texttt{sequandi} performs better on the experimentally-informed benchmark dataset (compare with Table~\ref{tab: 1}), which is the one where we are interested in optimal performance.
\begin{table}[ht]
\centering
\begin{tabular}{|c|c|c|c|c|c|c|}
\hline
Dataset & $\textrm{MAE}_{\alpha(t)}$ & $\textrm{MSLE}_{K(t)}$ & $\textrm{RMSE}_{CP}$ & $\textrm{JSC}_{CP}$ & $\textrm{MAE}_{\alpha}$ & $\textrm{MSLE}_{K}$ \\
\hline
Training & 0.228 & 0.122 & 1.947 & 0.293 & 0.258 & 0.191 \\
Restricted Train & 0.188 & 0.063 & 1.617 & 0.684 & 0.177 & 0.076 \\
Test & 0.194 & 0.064 & 1.421 & 0.699 & 0.187 & 0.075 \\
Benchmark & 0.208 & 0.028 & 1.509 & 0.483 & 0.217 & 0.031 \\
\hline
\end{tabular}
\caption{Performance of the tuned networks based on metrics of predictions in different datasets. $\textrm{MAE}_{\alpha(t)}$ and $\textrm{MSLE}_{K(t)}$ are the average of point-wise mean absolute error in $\alpha$ and mean squared logarithmic error in $K$, respectively, of all trajectories in each dataset. From the AnDi challenge, $\textrm{RMSE}_{CP}$ is the root mean squared error between true and predicted changepoint times that are true positives (TPs). $\textrm{JSC}_{CP}$ is the rate of these TPs (higher is better).  $\textrm{MAE}_{\alpha}$ and  $\textrm{MSLE}_{K}$ are the average errors of $\alpha$ and $K$ only in the ensemble of TP segment detections.}\label{tab: 2}
\end{table}

\section{Neural network changepoint detection signal}\label{app:NN-CP}
\begin{figure}[h!]
	\begin{center}
	\includegraphics[width=.92\textwidth]{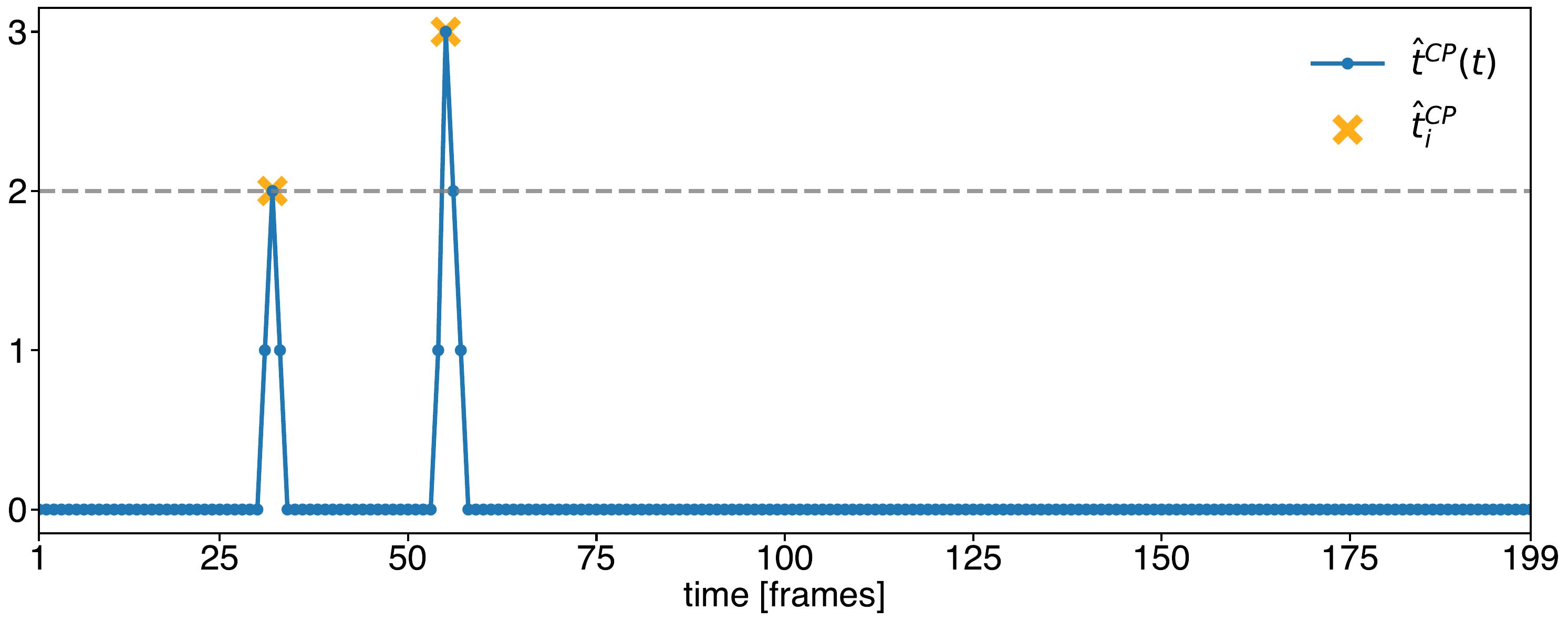}
	\end{center}
	\caption{{\bf Changepoint Detection from pooled neural network signal.} Pooled outputs of the two parallel layers of the neural network detecting and locating changepoints for the example trajectory shown in Fig~\ref{fig:fig1}.} 
	\label{fig:app_tCP}
\end{figure}
Figure~\ref{fig:app_tCP} shows an example of the pooled outputs from the two layers of the neural network devoted to detecting and locating changepoints for the trajectory shown in Fig~\ref{fig:fig1}.
As described in ~\ref{sec:tCP_NN}, we pool outputs from the neural network's changepoint detection predictions to compose a signal $\hat t^{\textrm{CP}}(t)$. The final changepoint predictions are extracted from this by finding the peaks of the signal above a threshold count of 2 (dashed gray line) and with a minimum time 3 between them.
\section{Changepoint algorithm optimisation}\label{app:CP}
In this Appendix, we provide additional details about our implementation of the CPDA developed in Ref.~\cite{li2015atp}. The algorithm consists of three main steps: detecting the presence of a CP, locating the CP and dealing with multiple CPs.\\
\begin{itemize}

\item\textit{Step 1: Detecting a changepoint} \\
The first step checks if there are changepoints in a trajectory segment. Here, we are using as input the prediction from our networks $\hat{\alpha}(t)$. A straight line $\alpha_{\text{fit}}(t)$ is fitted. The residual is computed as $d(t) = \hat{\alpha}(t) - \alpha_{\text{fit}}(t)$, followed by the cumulative sum $\text{CUMSUM}(t)$ of $d(t)$. The statistic $D_{\text{data}}$ is the difference between the maximum and minimum values of $\text{CUMSUM}(t)$.  
The significance of $D_{\text{data}}$ is tested by permuting $d(t)$ multiple times to generate a distribution of $\text{CUMSUM}(t)$ curves and $D_{\text{permute}}$ values. A confidence level (e.g., 95\%) sets the threshold $D_{\text{cutoff}}$. If $D_{\text{data}} \geq D_{\text{cutoff}}$, a changepoint is present. We discuss the relevance of the confidence level in~\ref{app:cl_CPDA}

\item\textit{Step 2: Locating the changepoint} \\ 
After confirming a changepoint, its position is found by testing each point $t$ in the trajectory. For each $t$, the segments before and after are fitted with straight lines, and the squared error (SE) is calculated as the sum of squared residuals. The $t$ with the smallest SE is the changepoint.

\item \textit{Step 3: Locating Multiple changepoints}\\  
To find multiple changepoints, the trajectory is split at each detected changepoint, and Steps 1 and 2 are repeated on each segment until no further changepoints are found.  

Each changepoint $\hat t^{\textrm{CP}}_{i}$ is then validated by re-analysing the segment from $\hat t^{\textrm{CP}}_{i-1}$ to $\hat t^{\textrm{CP}}_{i}$, applying Step 1’s hypothesis test and Step 2’s localisation to confirm its robustness.
\end{itemize}
\subsection{Confidence levels for the CPDA}\label{app:cl_CPDA}
The CPDA requires choosing the confidence level to make the decision about whether a CP is present in a segment. This is a crucial parameter determining the sensitivity of the changepoint detection. Setting the confidence too low results in many false positives (claiming a CP occurred when it did not), and setting it too high in many false negatives (missing CP that occurred).
\begin{figure}[h t]
	\begin{center}
	\includegraphics[width=\textwidth]{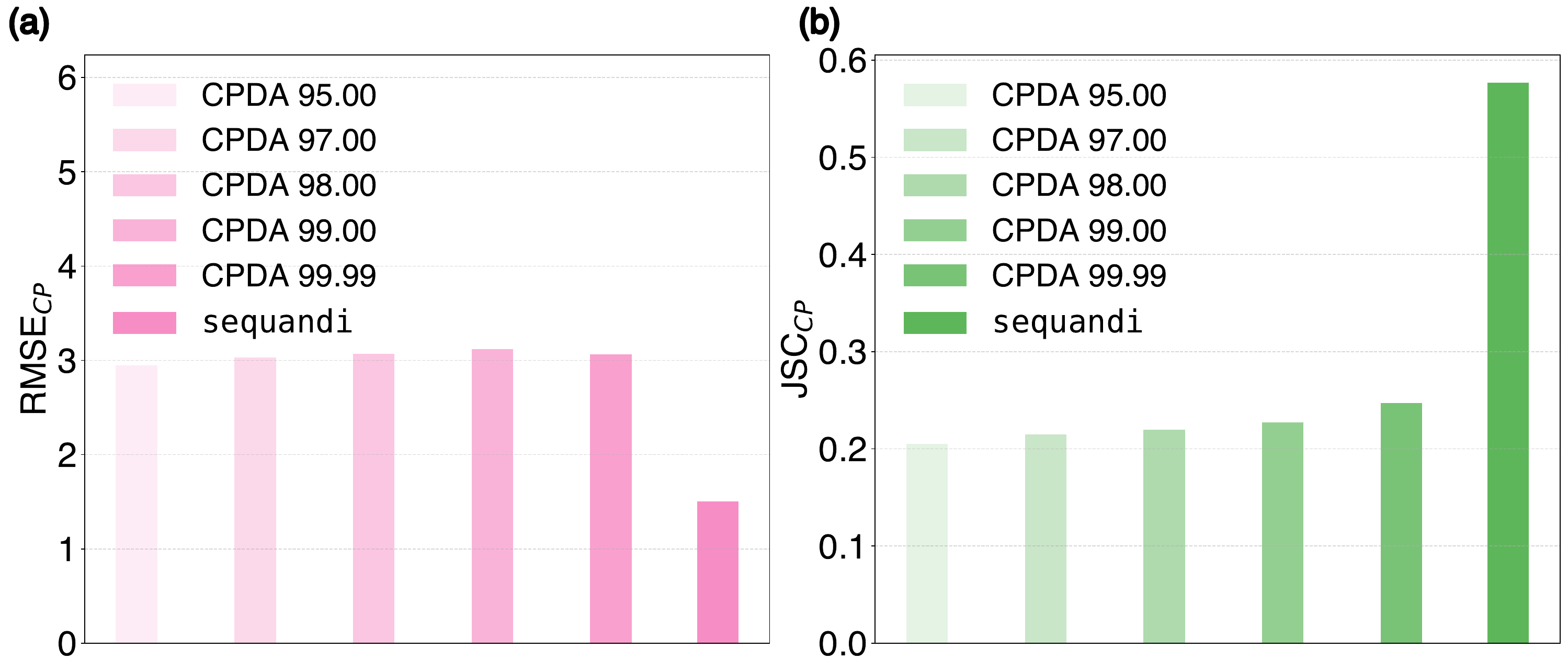}
	\end{center}
	\caption{{\bf Comparison of CPDA and $\mathtt{sequandi}$ Performance.} The confidence levels used are from ~\ref{app:cl_CPDA}. (a) shows the RMSE of different confidence levels for CPDA and Sequandi, while (b) presents the JSC performance under the same settings. Lower RMSE and higher JSC values indicate better performance.} 
	\label{fig:alg vs N.N.}
\end{figure}

For the AnDi 2 Benchmark dataset, we explore confidence levels of 95\%, 97\%, 98\%, 99\%, and 99.99\%, with a permutation number of 10,000. 
The ability of the CPDA to detect CPs increases with the confidence level, as shown in Fig.~\ref{fig:alg vs N.N.}(b). 
The precision of the detected CPs, quantified by the RMSE in Eq.~\eqref{eq:RMSE}, exhibits a non-monotonic trend, initially increasing before decreasing, as shown in Fig.~\ref{fig:alg vs N.N.}(a).\\
\begin{figure}[h t]
	\begin{center}
	\includegraphics[width=0.83\textwidth,trim={0.1cm 0.4cm 0.4cm 0},clip]{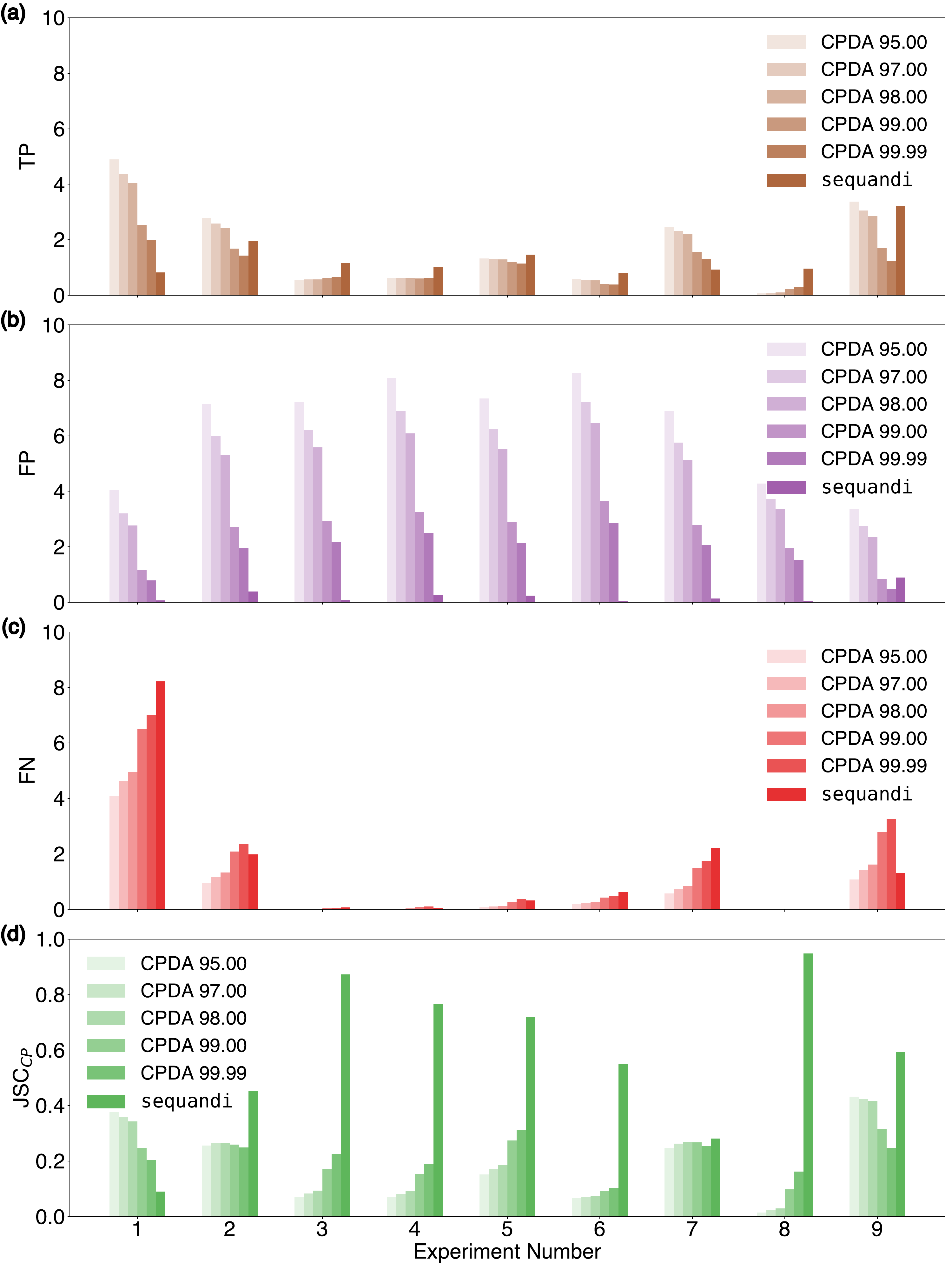}
	\end{center}
	\caption{{\bf Comparison of CPDA and $\texttt{sequandi}$ for each experiment} {\bf (a)}
    True positives (TP) of the changepoint detection normalised by the number of trajectories for each experiment in the benchmark dataset and different confidence levels of the CPDA. 
    {\bf (b)}
    False positives (FP) of the changepoint detection normalised by the number of trajectories for each experiment in the benchmark dataset and different confidence levels of the CPDA.
    {\bf (c)}
    False negatives (FN) of the changepoint detection normalised by the number of trajectories for each experiment in the benchmark dataset and different confidence levels of the CPDA.
    {\bf (d) }JSC for each experiment in the benchmark dataset and different confidence levels of the CPDA} 
	\label{fig:tpfpfn}
\end{figure}

It is interesting to consider the performance of the CPDA on each of the experiments of AnDi2024, which we report in Fig.~\ref{fig:tpfpfn}.
As expected, we see that a higher confidence level reduces the number of false positives [Fig.~\ref{fig:tpfpfn}(b)] but increases the number of false negatives
[Fig.~\ref{fig:tpfpfn}(c)]. The behaviour of the true positives  [Fig.~\ref{fig:tpfpfn}(a)] shows the opposite trend, with the exception of Experiment 3.
The JSC of the algorithm display a more intricate behaviour, which depends on the experiment.
For Experiments 1 and 9, which feature many changepoints, a low confidence interval returns the best JSC. For Experiments 2 and 7, there is an intermediate optimal confidence level. In the remaining experiments, which do not feature many changepoints, a high confidence level gives the best JSC.
\clearpage
\section{Visualising \texttt{sequandi} performances on example trajectories}
\subsection{Comparison with TA-MSD and Ref.~\cite{Bo2019}}\label{app:compare}
\begin{figure}[h ]
    \begin{center}
		\includegraphics[width=\textwidth]{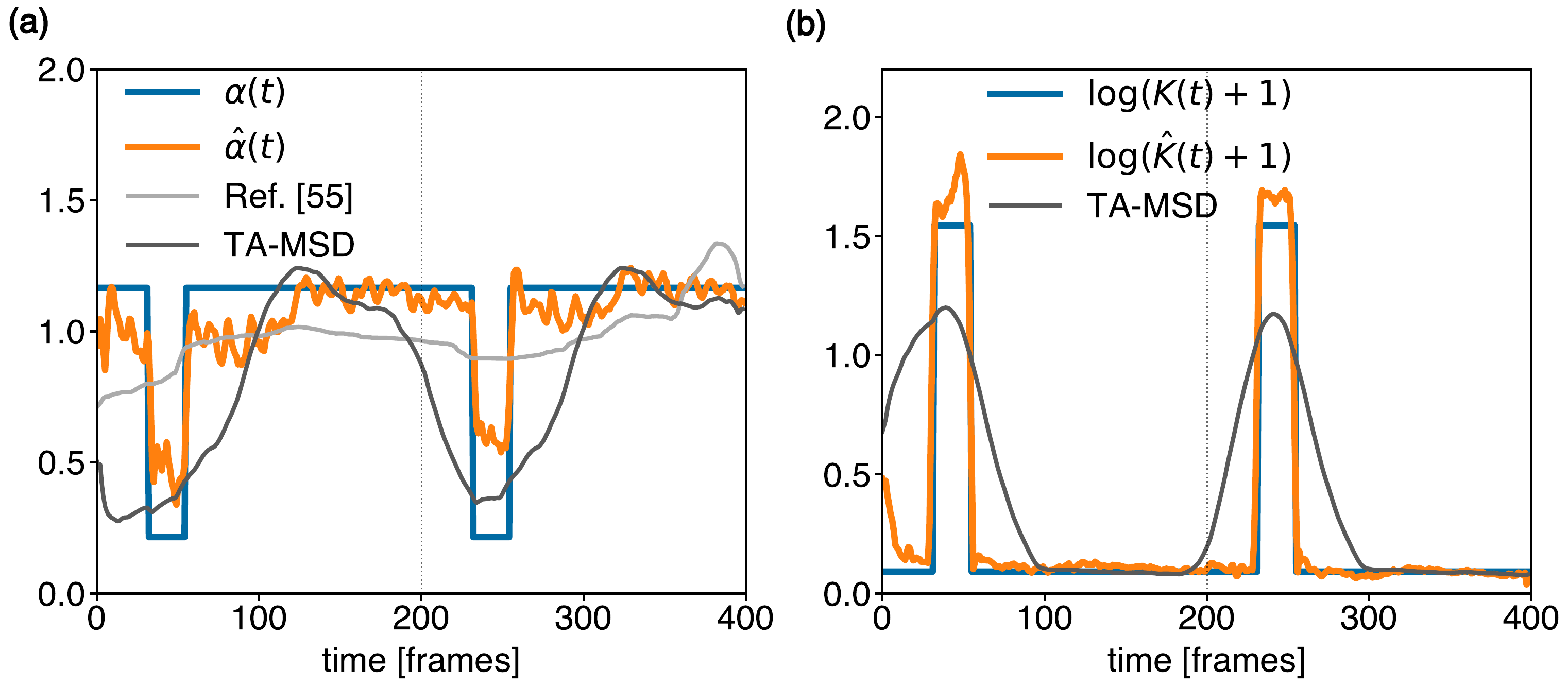}
	\end{center}
	\caption{{\bf Example of $\hat\alpha(t)$ and $\hat K(t)$ and ground truth for the trajectory of Fig.~\ref{fig:fig1} repeated twice.} 
	{\bf (a)} The orange line is \texttt{sequandi}'s point-wise prediction of the anomalous exponent $\hat\alpha(t)$, and the blue piece-wise constant line is the ground truth $\alpha(t)$. The grey line is the prediction with the method from  Ref.~\cite{Bo2019}, and the brown line is the TA-MSD prediction, computed with a sliding window of length 50.
    {\bf (b)} The orange line is based on \texttt{sequandi}'s point-wise prediction of the generalised diffusion coefficient $\hat K(t)$, and is 
    $\log(\hat{K}(t)+1)$.
    The blue piece-wise constant line is the ground truth $\log(K(t)+1)$, and the brown line is the TA-MSD prediction, computed with a sliding window of length 50. }
	\label{fig:fi3_app}
\end{figure}
A standard way to characterise anomalous diffusion from single trajectories is to compute the TA-MSD~\cite{Kepten2013,Kepten2015,Lanoiselee2018}.
If the process considered is ergodic~\cite{Jeon2010a},
which is the case for the fractional Brownian motion (fBm) trajectories considered here, plotting the TA-MSD in a log-log plot yields approximately a straight line where the slope is the anomalous diffusion exponent, and the offset is the generalised diffusion coefficient. We use the first 5 lag times to fit the straight line, as in Ref.~\cite{Bo2019}.
A more refined method, with excellent performance on fBm switching between different exponents (one switch), was derived in Ref.~\cite{Bo2019}. 
In Fig.~\ref{fig:fig3}, we compare \texttt{sequandi} to these traditional methods for the challenging case of multiple switches. Since the method of Ref.~\cite{Bo2019} requires a segment of at least 256 points, we apply it to two consecutive copies of the trajectory of Fig.~\ref{fig:fig1}, which gives a trajectory containing 400 data points, switching between different diffusive behaviours as shown in Fig.~\ref{fig:fi3_app}. In Fig.~\ref{fig:fig3} in the main text, we report the results on the first 200 points of the trajectory.  
 In Fig.~\ref{fig:fi3_app} in this Appendix, we report the performance on the duplicated trajectory, containing 400 points.
 Even though \texttt{sequandi} was trained on trajectories of length 200, it can easily generalise to longer trajectories and greatly outperform the traditional methods.
 For the original first 200 points, \texttt{sequandi} tracks the ground truth with an $\textrm{MAE}_{\alpha(t)}=0.134$  and $\textrm{MSLE}_{K(t)}=0.016$. Using the repeated trajectory of 400 frames, \texttt{sequandi} achieves a comparable and even better performance, $\textrm{MAE}_{\alpha(t)}=0.116$  and $\textrm{MSLE}_{K(t)}=0.011$. The RNN from Ref.~\cite{Bo2019} has $\textrm{MAE}_{\alpha(t)}=0.248$, and the TA-MSD curve has  $\textrm{MAE}_{\alpha(t)}=0.280$ and $\textrm{MSLE}_{K(t)}=0.386$.

\subsection{Additional example trajectories}\label{app:more}
\begin{figure}[h]
	\begin{center}
	\includegraphics[width=\textwidth]{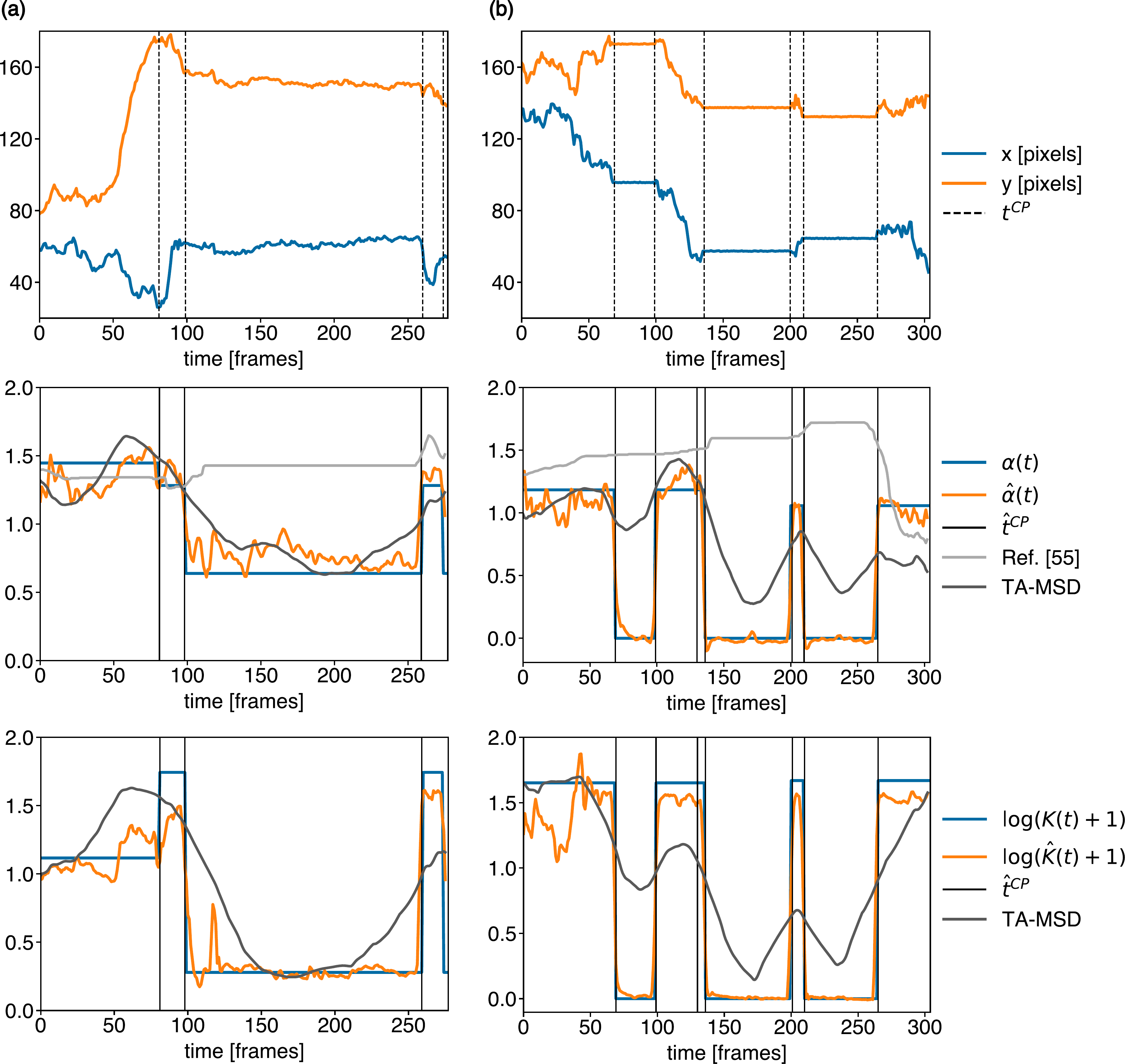}
	\end{center}
	\caption{{\bf Inference of diffusion parameters and segments on two additional trajectories (a) and (b)} Top row: $x$ and $y$ coordinate tracks of the diffusive trajectories, with the ground truth changepoint times $t^\textrm{CP}$ denoted by the dashed vertical line. Middle row: anomalous diffusion exponent evolution with ground truth $\alpha (t)$ in blue, and point-wise predictions $\hat\alpha (t)$ in orange. Predicted changepoint times $\hat t^\textrm{CP}$ given by the solid black vertical lines. The grey line is the prediction with the method from  Ref.~\cite{Bo2019}, and the brown line is the TA-MSD prediction, computed with a sliding window of length 50. \textit{Bottom row}: likewise for the generalised diffusion coefficient $K(t)$. For the \texttt{sequandi} predictions in {\bf (a)}, $\textrm{MAE}_{\alpha(t)}=0.132$ and $\textrm{MSLE}_{K(t)}=0.040$. For {\bf (b)}, $\textrm{MAE}_{\alpha(t)}=0.079$ and $\textrm{MSLE}_{K(t)}=0.045$.} 
	\label{fig:extraTraj}
\end{figure}

To complement the main text, in this appendix we provide more examples of point-wise and CP location inference on individual trajectories, generated from the same ensemble as the test dataset in Fig.~\ref{fig:app_restricted}. The trajectories are longer than 256 frames, to accommodate the use of the Ref.~\cite{Bo2019} network without needing to repeat them. The $x$ and $y$ trajectories of the two new trajectories, {\bf (a)} and {\bf (b)} in Fig.~\ref{fig:extraTraj}, are plotted in the top row. {\bf (a)} features an intermediate and long duration segment of different state parameters, with short switches in between. {\bf (b)} has rapid hopping throughout, visiting a trapped, immobile state.
\texttt{sequandi} manages to track large jumps in $K$, with inference of both parameters improving with longer segment duration. However, we highlight once more that these are very frequent switching of diffusion states, and trajectories overall have a short duration.  \texttt{sequandi} detects accurately the changepoints in both trajectories and wrongly identifies a false positive in trajectory {\bf (b)}.
Comparing \texttt{sequandi} with the network of \cite{Bo2019}, which struggles on these short segments and the
 TA-MSD computed with sliding windows of total length 50, which coarsely tracks the varying diffusion parameters confirms the great improvement provided by 
  \texttt{sequandi} for trajectories featuring frequent switches.

\clearpage
\section{Benchmark dataset experiment-wise performance}

In this Appendix, we report the performance of the segment-wise predictions on the benchmark dataset discussed in Section~\ref{sec:bench} for each experiment.
\begin{figure}[h]
	\begin{center}
	\includegraphics[width=\textwidth]{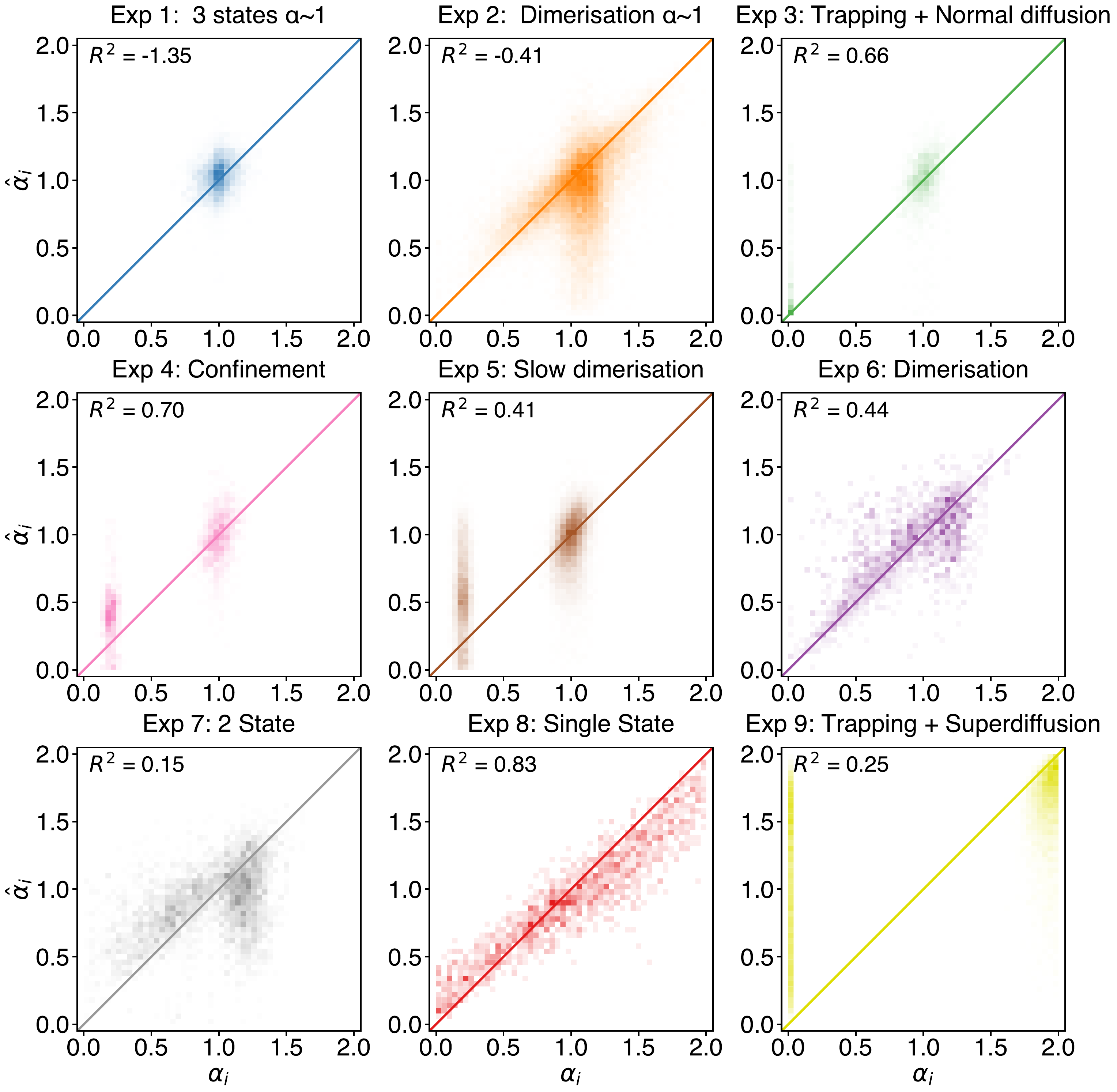}
	\end{center}
	\caption{{\bf Ground truth vs. predicted anomalous diffusion exponent.} For each experiment in the benchmark dataset, a density map of the predicted segment-wise $\hat \alpha_i$ and their respective ground truths.} 
	\label{fig:app_alpha}
\end{figure}
\clearpage
\begin{figure}[h]
	\begin{center}
	\includegraphics[width=\textwidth]{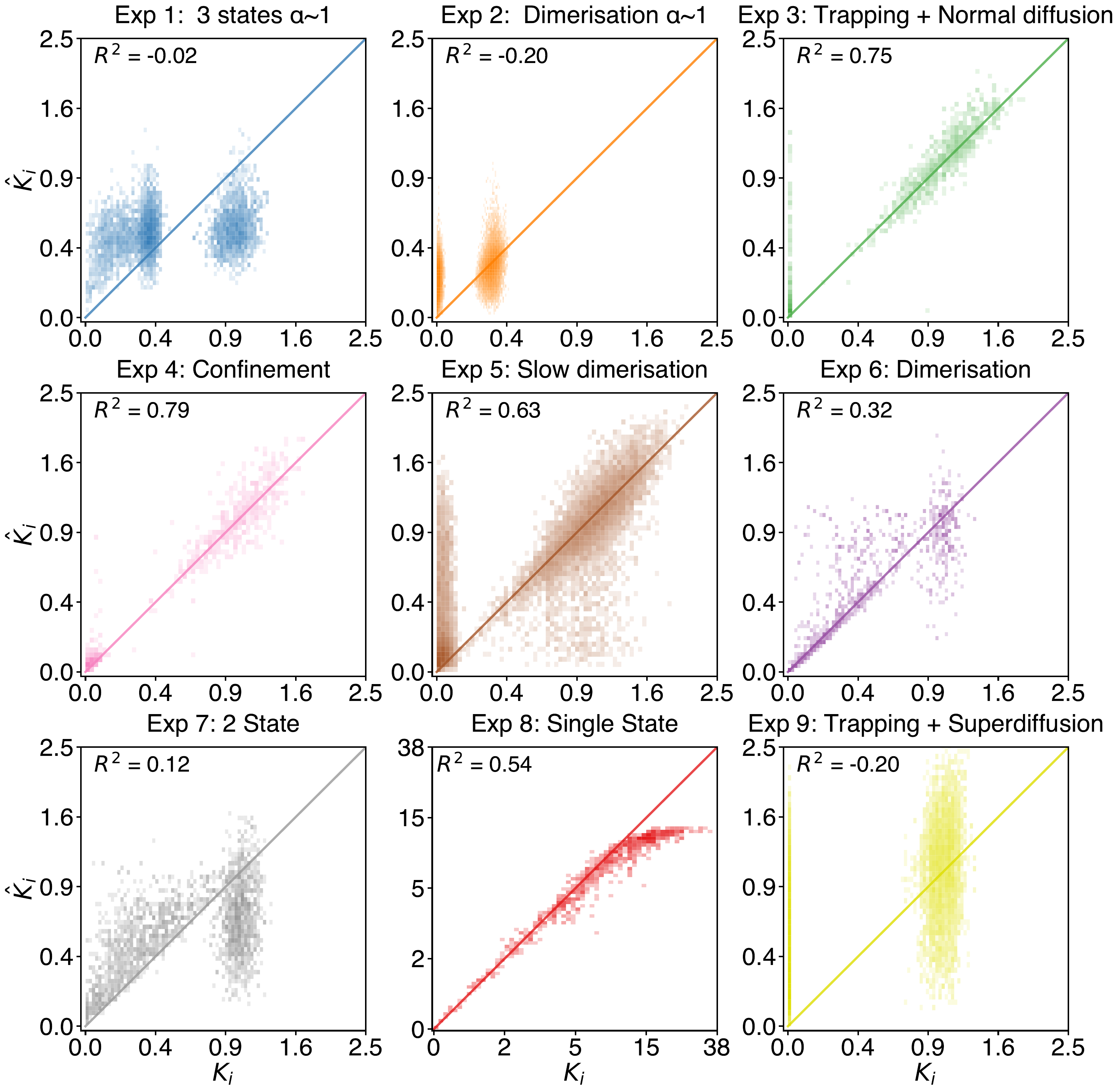}
	\end{center}
	\caption{{\bf Ground truth vs. predicted generalised diffusion coefficient.} For each experiment in the benchmark dataset, a density map of the predicted segment-wise $\hat K_i$ and their respective ground truths. The x and y axis scales are logarithmic, and the colourisation of the density map is with a logarithmic normalisation due to the more dilute grid.} 
	\label{fig:app_K}
\end{figure}

\bibliography{sequandi}
\end{document}